\documentclass{article} 
\usepackage{graphicx}
\usepackage{color}
\usepackage{amsmath}
\usepackage{amsfonts}
\usepackage[normal]{subfigure}
\usepackage{a4wide}
\usepackage{enumitem}
\usepackage{authblk}
\usepackage{amsfonts}

\usepackage{amsmath}
\usepackage{amssymb}
\usepackage{latexsym}
\usepackage[utf8]{inputenc}
\usepackage{engtlc}
\usepackage{graphicx}
\usepackage{amsthm}
\usepackage{engtlc}
\usepackage{listings}
\usepackage{xcolor}
\usepackage{bbm}
\usepackage{placeins}
\usepackage{booktabs}
\usepackage{setspace}
\usepackage{multicol}
\usepackage{float}
\usepackage{xcolor}

\providecommand{\keywords}[1]{\textbf{Keywords:} #1}
\newtheorem{assumption}{Assumption}
\newtheorem{remark}{Remark}
\newcommand{\interior}[1]{%
	{\kern0pt#1}^{\mathrm{o}}%
}

\date{}
\begin{document}

\title{A mathematical study of the influence of hypoxia and acidity on the evolutionary dynamics of cancer\thanks{This work was supported by the MIUR grant ``Dipartimenti di Eccellenza 2018-2022''.
}
}

\author{Giada Fiandaca \thanks{Department of Mathematical Sciences ``G. L. Lagrange'', Politecnico di Torino, Corso Duca degli Abruzzi, 24, Torino, Italy} \quad \quad \quad Marcello Delitala \thanks{
            Department of Mathematical Sciences ``G. L. Lagrange'', Politecnico di Torino, Corso Duca degli Abruzzi, 24, Torino, Italy} \quad \quad \quad  Tommaso Lorenzi \thanks{
             Department of Mathematical Sciences ``G. L. Lagrange'', Politecnico di Torino, Corso Duca degli Abruzzi, 24, Torino, Italy}}
\maketitle
\begin{abstract}
Hypoxia and acidity act as environmental stressors promoting selection for cancer cells with a more aggressive phenotype. As a result, a deeper theoretical understanding of the spatio-temporal processes that drive the adaptation of tumour cells to hypoxic and acidic microenvironments may open up new avenues of research in oncology and cancer treatment. We present a mathematical model to study the influence of hypoxia and acidity on the evolutionary dynamics of cancer cells in vascularised tumours. The model is formulated as a system of partial integro-differential equations that describe the phenotypic evolution of cancer cells in response to dynamic variations in the spatial distribution of three abiotic factors that are key players in tumour metabolism: oxygen, glucose and lactate. The results of numerical simulations of a calibrated version of the model based on real data recapitulate the eco-evolutionary spatial dynamics of tumour cells and their adaptation to hypoxic and acidic microenvironments. Moreover, such results demonstrate how nonlinear interactions between tumour cells and abiotic factors can lead to the formation of environmental gradients which select for cells with phenotypic characteristics that vary with distance from intra-tumour blood vessels, thus promoting the emergence of intra-tumour phenotypic heterogeneity. Finally, our theoretical findings reconcile the conclusions of earlier studies by showing that the order in which resistance to hypoxia and resistance to acidity arise in tumours depend on the ways in which oxygen and lactate act as environmental stressors in the evolutionary dynamics of cancer cells.\\
\\
\end{abstract}
\section{Introduction}
\label{intro}
Cancer is a dynamic disease, the characteristics of which are constantly evolving. This is reflected in the fact that the genotypic and phenotypic properties of cancer cells may change across space and time within the same tumour,  and the dynamics of tumours with the same histological features are still likely to vary across patients. Moreover, since the same cancer clones may arise through different evolutionary pathways, the fact that two tumours have a similar clonal composition at a given point in time does not necessarily indicate that they share similar evolutionary histories, and does not rule out the possibility that their future evolution will diverge significantly~\cite{maley2017classifying}. These sources of variability within and between tumours provide the substrate for the emergence and development of intra- and inter-tumour heterogeneity, which are major obstacles to cancer eradication~\cite{gillies2012evolutionary,marusyk2012intra}.

Clinical evidence suggests that cancer cells and the tumour microenvironment mutually shape each other~\cite{Gallaher270900}. This supports the idea that tumours can be seen as evolving ecosystems where cancer cells with different phenotypic characteristics proliferate, die, undergo genotypic and phenotypic changes, and compete for space and resources under the selective pressure exerted by the various components of the tumour microenvironment~\cite{gallaher2013evolution,gay2016tumour,ibrahim2017defining,korolev2014turning,lloyd2016darwinian,loeb2001mutator,merlo2006cancer,michor2010origins,villa2021modeling,vander2020reconciling}. In this light, intra-tumour phenotypic heterogeneity can be regarded as the outcome of an eco-evolutionary process in which spatial variability of the concentration of abiotic factors (\emph{i.e.} substrates and metabolites) across the tumour supports the formation of distinct ecological niches whereby cells with different phenotypic characteristics may be selected~\cite{casciari1992variations,gatenby2007cellular,hockel2001tumor}.

In normal tissues, cells produce the energy required to sustain their proliferation via oxidative phosphorylation (\emph{i.e.} they rely on oxygen as their primary source of energy) and turn to glycolysis only when oxygen is scarce. In tumours, the presence of hypoxic regions (\textit{i.e.} regions where the oxygen levels are below the physiological ones) induces cells to transiently switch to a glycolytic metabolic phenotype (\emph{i.e.} to rely on glucose as their primary source of energy)~\cite{Treatment}. Cancer cells eventually acquire such a glycolytic phenotype and express it also in aerobic conditions, leading to the so-called Warburg effect~\cite{Kim8927}. The interplay between the high glycolytic rate of cancer cells and low perfusion in tumours brings about accumulation of lactate (\emph{i.e.} a waste product of glycolysis), which causes acidity levels in the tumour microenvironment to rise (\emph{i.e.} the pH level drops)~\cite{tang2012functional}. 

Since hypoxia and acidity act as environmental stressors promoting selection for cancer cells with a more aggressive phenotype~\cite{Robertson-Tessi1567,tang2012functional}, an in-depth theoretical understanding of the spatio-temporal processes that drive the adaptation of tumour cells to hypoxic and acidic microenvironments may open up new avenues of research in oncology and cancer treatment~\cite{therapy1}. \textcolor{black}{In this regard, mathematical models can be an important source of support to cancer research, as they enable extrapolation beyond scenarios which can be investigated through experiments and may reveal emergent phenomena that would otherwise remain unobserved~\cite{anderson2008integrative,byrne2010dissecting,chaplain2020multiscale,chisholm2016cell,eastman2020effects,hamis2020,poleszczuk2015evolution}. For instance, in their pioneering paper~\cite{Gatenby5745}, Gatenby and Gawlinski used a reaction-diffusion system to explore how nonlinear interactions between cancer cells and abiotic components of the tumour microenvironment may shape tumour growth. The Gatenby-Gawlinski model has recently been extended in~\cite{strobl2020mix}, in order to take into account the presence of cells with different phenotypic characteristics within the tumour. Hybrid cellular automaton models have been employed to study the impact of hypoxia and acidity on tumour growth and invasion~\cite{anderson2006tumor,anderson2009microenvironment,gatenby2007cellular,hatzikirou2012go,kim2018role,Robertson-Tessi1567}. A mechanical model of tumour growth whereby cells are allowed to switch between aerobic and anaerobic metabolism was presented in~\cite{ASTANIN2009578}. Integro-differential equations and partial integro-differential have been used in~\cite{ardavseva2019mathematical,chaplain2019evolutionary,LORENZI2018101,lorz2015modeling,villa2021modeling} to investigate the ecological role of hypoxia in the development of intra-tumour phenotypic heterogeneity.} 

In this paper, we complement these earlier studies by presenting a mathematical model to study the influence of hypoxia and acidity on the evolutionary dynamics of cancer cells in vascularised tumours. The model comprises a system of partial integro-differential equations that describe the phenotypic evolution of cancer cells in response to dynamic variations in the spatial distribution of three abiotic factors that are key players in tumour metabolism: oxygen, glucose and lactate. 

The remainder of the paper is organised as follows. In Section~\ref{Sec2}, we present the model equations and the underlying modelling assumptions. In Section~\ref{Sec3}, we summarise the main results of numerical simulations of the model and discuss their biological implications. Section~\ref{Sec4} concludes the paper and provides a brief overview of possible research perspectives.
\section{Model description}
\label{Sec2}
\begin{figure}[htp!]
	\centering
	\includegraphics[scale=0.8]{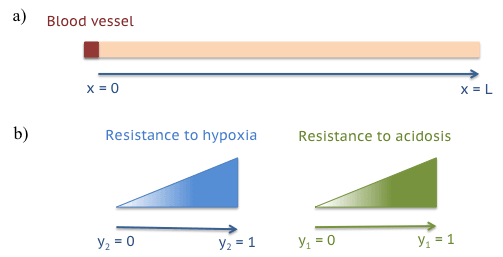}
	\caption{\footnotesize{{\bf a)} Schematic of the spatial domain defined as a one-dimensional region of vascularised tissue of length ${\rm L}$. A blood vessel is assumed to be present at $x=0$. {\bf b)} Schematics illustrating the relationships between the values of the variables $y_2$ and $y_1$ modelling the phenotypic state of tumour cells and the levels of resistance to hypoxia and acidosis.}}
	\label{figure1}
\end{figure}

We consider a one-dimensional region of vascularised tissue of length ${\rm L} > 0$. We describe the spatial position of every tumour cell in the tissue region by a scalar variable $x \in [0,{\rm L}]$ and we assume a blood vessel to be present at $x=0$ ({\it cf.} the schematic in Figure~\ref{figure1}a). Moreover, building upon the modelling framework developed in~\cite{chaplain2019evolutionary,LORENZI2018101,lorz2015modeling,villa2021modeling}, we model the phenotypic state of every cell by a vector ${\bf y} = (y_1,y_2) \in [0,1]^2$ ({\it cf.} the schematics in Figure~\ref{figure1}b). Here, $y_1 \in [0,1]$ represents the normalised level of expression of an acidity-resistant gene (\emph{e.g.} the LAMP2 gene), while $y_2 \in [0,1]$ represents the normalised level of expression of a hypoxia-resistant gene (\emph{e.g.} the GLUT-1 gene)~\cite{Damaghi2015ChronicAI,gatenby2007cellular}.

We describe the phenotypic distribution of tumour cells at position $x$ and time $t \in [0,{\rm T}]$, with ${\rm T}>0$, by means of the local population density function $n(t,x,{\bf y})$ (\emph{i.e.} the local phenotypic distribution of tumour cells). We define the cell density $\rho(t,x)$, the local mean level of expression of the acidity-resistant gene $\mu_1(t,x)$ and the local mean level of expression of the hypoxia-resistant gene $\mu_2(t,x)$ as
\begin{equation}
\rho(t,x):=\int_{[0,1]^2} n(t,x,{\bf y}) \,{\rm d}{\bf y}, \quad 
\mu_i(t,x):=\frac{1}{\rho(t,x)}\int_{[0,1]^2} y_i\,n(t,x,{\bf y})\,{\rm d}{\bf y}
\label{def:rhomu}
\end{equation}
for $i=1,2$. Moreover, we define the phenotypic distribution of tumour cells across the whole tissue region $f(t,{\bf y})$ as the mean value of $n(t,x,{\bf y})$ on the interval $[0,{\rm L}]$, \emph{i.e.} 
\begin{equation}
f(t,{\bf y}) := \frac{1}{{\rm L}} \, \int_0^{{\rm L}} n(t,x,{\bf y}) \, {\rm d}x.
\label{def:f}
\end{equation}	
Similarly, we define the levels of expression of the acidity-resistant gene and the hypoxia-resistant gene across the whole tissue region as the mean values of $\mu_1(t,x)$ and $\mu_2(t,x)$ on the interval $[0,{\rm L}]$, respectively, \emph{i.e.} 
\begin{equation}
\nu_1(t) := \frac{1}{{\rm L}} \, \int_0^{{\rm L}} \mu_1(t,x) \, {\rm d}x \quad \text{and} \quad \nu_2(t) := \frac{1}{{\rm L}} \, \int_0^{{\rm L}} \mu_2(t,x) \, {\rm d}x.
\label{def:nu}
\end{equation}	

The local concentrations of oxygen, glucose and lactate at position $x$ and time $t$ are denoted by $S_o(t,x)$, $S_g(t,x)$ and $S_l(t,x)$, respectively.
\medskip
\subsection{Dynamics of tumour cells}
We describe the dynamics of tumour cells through the following balance equation for the local population density function $n(t,x,{\bf y})$
\begin{equation}
\frac{\partial n}{\partial t} =  \underbrace{\beta_n \frac{\partial^2 n}{\partial x^2}}_{\substack{\text{undirected, \,random}\\\text{cell movement}}}+ \underbrace{\theta \, \Delta_{{\bf y}} n}_{\substack{\text{spontaneous} \\ \text{phenotypic changes}}} + \;\, \underbrace{ R\,(S_o,S_g,S_l,\rho,{\bf y})\,n}_{\text{proliferation and death}},
\label{main}
\end{equation}
with $(t,x,{\bf y}) \in (0,{\rm T}] \times (0,{\rm L}) \times (0,1)^2$, subject to suitable initial conditions. We complement~\eqref{main} with zero-flux boundary conditions at $x=0$ and $x={\rm L}$ (\emph{i.e.} we assume that cells cannot leave the tissue region) and zero-flux boundary conditions on the boundary of the square $[0,1]^2$ (\emph{i.e.} we assume that cells cannot have normalised levels of gene expression smaller than $0$ or larger than $1$). 

The first term on the right-hand side of the partial integro-differential equation~\eqref{main} models the effect of undirected, random movement, which is described through Fick's first law of diffusion with diffusivity $\beta_n>0$. 

\textcolor{black}{The second term on the right-hand side of the partial integro-differential equation~\eqref{main} models the effect of heritable, spontaneous phenotypic changes, which occur at rate $\theta>0$. Similar diffusion terms have been used in a number of previous papers to model the effect of spontaneous phenotypic changes~\cite{alfaro2019evolutionary,almeida2019evolution,ardavseva2020evolutionary,bouin2012invasion,chisholm2015emergence,figueroa2018long,genieys2006pattern,Lorenzi2016,mirrahimi2020evolution,perthame2007concentration} and can be obtained as the deterministic continuum limit of corresponding stochastic individual-based models in the asymptotic regime of large numbers of individuals and small phenotypic changes~\cite{champagnat2006unifying,chisholm2016disco}.}

The function $R(S_o,S_g,S_l,\rho,{\bf y})$ represents the fitness of tumour cells in the phenotypic state ${\bf y}$ at position $x$ and time $t$ under the environmental conditions given by the concentrations of abiotic factors $S_o(t,x)$, $S_g(t,x)$ and $S_l(t,x)$, and the cell density $\rho(t,x)$ (\emph{i.e.} $R(S_o,S_g,S_l,\rho,{\bf y})$ is the phenotypic fitness landscape of the tumour). We use the following definition
\begin{equation}
\label{eq:defR}
R(S_o,S_g,S_l,\rho,{\bf y}) \; := \underbrace{P(S_o,S_g, y_2)}_{\substack{\text{proliferation and}\\\text{death due to} \\\text{oxygen-driven selection}}} - \underbrace{D(S_l,y_1)}_{\substack{\text{death due to}\\\text{lactate-driven selection}}} - \underbrace{d(\rho).}_{\substack{\text{death due to}\\\text{competition} \\\text{for space}}}
\end{equation}
Here, the function $P(S_o,S_g, y_2)$ is the rate at which cells with level of expression $y_2$ of the hypoxia-resistant gene proliferate via oxidative phosphorylation and glycolysis, and die due to oxygen-driven selection (\emph{i.e.} $P(S_o,S_g, y_2)$ is a net proliferation rate). The function $D(S_l,y_1)$ is the rate at which cells with level of expression $y_1$ of the acidity-resistant gene die due to lactate-driven selection. The function $d(\rho)$ models the rate of cell death due to competition for space associated with saturation of the cell density.

\subsubsection{Modelling oxygen-driven selection}
Based on the theoretical results and experimental data presented in~\cite{korolev2014turning,Treatment}, we focus on a scenario corresponding to the following biological assumptions.
\begin{assumption}
	\label{Ass0}
	There exist two threshold levels of the oxygen concentration $O_M>O_m>0$ such that the environment surrounding the cells is: hypoxic if $S_o \leq O_m$; moderately oxygenated if $O_m < S_o < O_M$;  normoxic (i.e. well oxygenated) if $S_o \geq O_M$. 
\end{assumption}

\begin{assumption}
	\label{Ass1}
	Cells proliferate at a rate that depends on the concentrations of oxygen and glucose. Moreover, the trade-off between increase in cell death associated with sensitivity to hypoxia and decrease in cell proliferation associated with acquisition of resistance to hypoxia results in the existence of a level of expression of the hypoxia-resistant gene which is the fittest in that: a lower level of gene expression would correlate with a lower resistance to hypoxia, and thus a higher death rate; a higher level of gene expression would correlate with a larger fitness cost, and thus  a lower proliferation rate. Cells with levels of gene expression that are closer to the fittest one are more likely to survive than the others. Hence, the farther the gene expression level of a cell is from the fittest one, the more likely is that the cell will die due to a form of oxygen-driven selection.
\end{assumption}

\begin{assumption}
	\label{Ass2}
	In normoxic environments (i.e. when $S_o \geq O_M$), the energy required for cell proliferation is produced via oxidative phosphorylation and the cell proliferation rate is a monotonically increasing function of the concentration of oxygen. In hypoxic environments (i.e. when $S_o \leq O_m$), the energy required for cell proliferation is produced via glycolysis and the cell proliferation rate is a monotonically increasing function of the concentration of glucose. In moderately-oxygenated environments (i.e. when $O_m < S_o < O_M$), the energy required for cell proliferation is produced via both oxidative phosphorylation and glycolysis. Moreover, the cell proliferation rate is a monotonically increasing function of the concentrations of oxygen and glucose, and lower values of the oxygen concentration correlate with a greater tendency of the cells to proliferate via glycolysis.
\end{assumption}

\begin{assumption}
	\label{Ass3}
	The fittest level of expression of the hypoxia-resistant gene (i.e. the gene associated with the variable $y_2$) may vary with the oxygen concentration. In particular: in normoxic environments (i.e. when $S_o \geq O_M$), the fittest level of gene expression is the minimal one (i.e. $y_2=0$); in hypoxic environments (i.e. when $S_o \leq O_m$) the fittest level of gene expression is the maximal one (i.e. $y_2=1$); in moderately-oxygenated environments (i.e. when $O_m < S_o < O_M$), the fittest level of gene expression is a monotonically decreasing function of the oxygen concentration (i.e. it decreases from $y_2=1$ to $y_2=0$ when the oxygen concentration increases).
\end{assumption}

Under Assumptions~\ref{Ass0} and~\ref{Ass1}, we define the net proliferation rate $P(S_o,S_g, y_2)$ as
\begin{equation}
P(S_o,S_g, y_2) \; :=  \underbrace{p_o(S_o)}_{\substack{\text{proliferation via}\\\text{oxidative phosphorylation}}} + \underbrace{p_g(S_o,S_g)}_{\substack{\text{proliferation via}\\\text{glycolysis}}} -  \underbrace{\eta_o \, \big(y_2-\varphi_o(S_o)\big)^2}_{\substack{\text{death due to} \\\text{oxygen-driven selection}}}.
\label{proliferation}
\end{equation}
In~\eqref{proliferation}, the function $p_o(S_o)$ models the rate of cell proliferation via oxidative phosphorylation, while the function $p_g(S_o,S_g)$ models the rate of cell proliferation via glycolysis. Furthermore, the third term in the definition given by~\eqref{proliferation} is the rate of death induced by oxygen-driven selection. Here, the parameter $\eta_o>0$ is a selection gradient that quantifies the intensity of oxygen-driven selection and the function $\varphi_o(S_o)$ is the fittest level of expression of the hypoxia-resistant gene under the environmental conditions given by the oxygen concentration $S_o$. 

\begin{remark}	
	\label{remark}
	In~\eqref{proliferation}, the distance between $y_2$ and $\varphi_o(S_o)$ is computed as $\big(y_2-\varphi_o(S_o)\big)^2$. Alternatively, one could compute such a distance as $\big|y_2-\varphi_o(S_o)\big|$. However, we have chosen $\big(y_2-\varphi_o(S_o)\big)^2$ over $\big|y_2-\varphi_o(S_o)\big|$ because, as discussed in~\cite{ardavseva2019mathematical,Lorenzi2016}, it leads to a smoother fitness function which is closer to the approximate fitness landscapes which can be inferred from experimental data through regression techniques. 
\end{remark}	

Under Assumptions~\ref{Ass2} and~\ref{Ass3}, we use the definitions of the functions $p_o(S_o)$, $p_g(S_o,S_g)$ and $\varphi_o(S_o)$ given hereafter
\begin{equation}
p_o(S_o) := \dfrac{\gamma_o\,S_o}{\alpha_o+S_o}\,w(S_o), \qquad p_g(S_o,S_g) := \dfrac{\gamma_g\,S_g}{\alpha_g+S_g}\,\big(1-w(S_o)\big),
\label{def:popg}
\end{equation}
with
\begin{equation}
w(S_o):=
\begin{cases}
1 \qquad  \qquad  \quad  \quad \quad \quad S_o           \geq O_M\\
\\
\displaystyle
1 - \frac{O_M-S_o}{O_M-O_m}  \quad  \quad O_m < S_o < O_M\\
\\
\displaystyle
0 \qquad  \qquad \quad  \quad \quad \quad S_o \leq O_m
\end{cases}	
\label{wo_So}
\end{equation}
and
\begin{equation}
\varphi_o(S_o):=
\begin{cases}
0 \qquad  \qquad  \quad  \quad S_o           \geq O_M\\
\\
\displaystyle
\frac{O_M-S_o}{O_M-O_m}  \quad  \quad O_m < S_o < O_M\\
\\
\displaystyle
1 \qquad  \qquad \quad  \quad S_o \leq O_m.
\end{cases}	
\label{phio_So}
\end{equation}

In~\eqref{def:popg}, the parameters $\gamma_o>0$ and $\gamma_g>0$ model the maximum rates of cell proliferation via oxidative phosphorylation and glycolysis, respectively. The parameters $\alpha_o>0$ and  $\alpha_g>0$ are the Michaelis-Menten constants of oxygen and glucose. The weight function $w(S_o)$ defined via~\eqref{wo_So} ensures that Assumption~\ref{Ass2} is satisfied, while definition~\eqref{phio_So} of $\varphi_o(S_o)$ is such that Assumption~\ref{Ass3} is satisfied ({\it cf.} the plot in Figure~\ref{figure2}a).

\subsubsection{Modelling lactate-driven selection}
Based on theoretical results and experimental data presented in~\cite{Robertson-Tessi1567}, we focus on a scenario corresponding to the following biological assumptions.

\begin{assumption}
	\label{Ass4}
	There exist two threshold levels of the lactate concentration $L_M>L_m>0$ such that the environment surrounding the cells is: mildly acidic if $S_l \leq L_m$; moderately acidic if $L_m < S_l < L_M$;  highly acidic if $S_l \geq L_M$.
\end{assumption}

\begin{assumption}
	\label{Ass5}
	Cells die at a rate that depends on the concentration of lactate. Moreover, the trade-off between increase in cell death associated with sensitivity to acidity and decrease in cell proliferation associated with acquisition of resistance to acidity results in the existence of a level of expression of the acidity-resistant gene which is the fittest in that: a lower level of gene expression would correlate with a lower resistance to acidity, and thus a higher death rate; a higher level of gene expression would correlate with a larger fitness cost, and thus a lower proliferation rate. Cells with levels of gene expression that are closer to the fittest one are more likely to survive than the others. Hence, the farther the gene expression level of a cell is from the fittest one, the more likely is that the cell will die due to a form of lactate-driven selection.
\end{assumption}

\begin{assumption}
	\label{Ass6}
	The fittest level of expression of the acidity-resistant gene (i.e. the gene associated with the variable $y_1$) may vary with the lactate concentration. In particular: in mildly-acidic environments (i.e. when $S_l \leq L_m$), the fittest level of gene expression is the minimal one (i.e. $y_1=0$); in highly-acidic environments (i.e. when $S_l \geq L_M$) the fittest level of gene expression is the maximal one (i.e. $y_1=1$); in moderately-acidic environments (i.e. when $L_m < S_l < L_M$), the fittest level of gene expression is a monotonically increasing function of the lactate concentration (i.e. it increases from $y_1=0$ to $y_1=1$ when the lactate concentration increases).
\end{assumption}

\begin{figure}[htp!]
	\centering
	\includegraphics[width=1\textwidth]{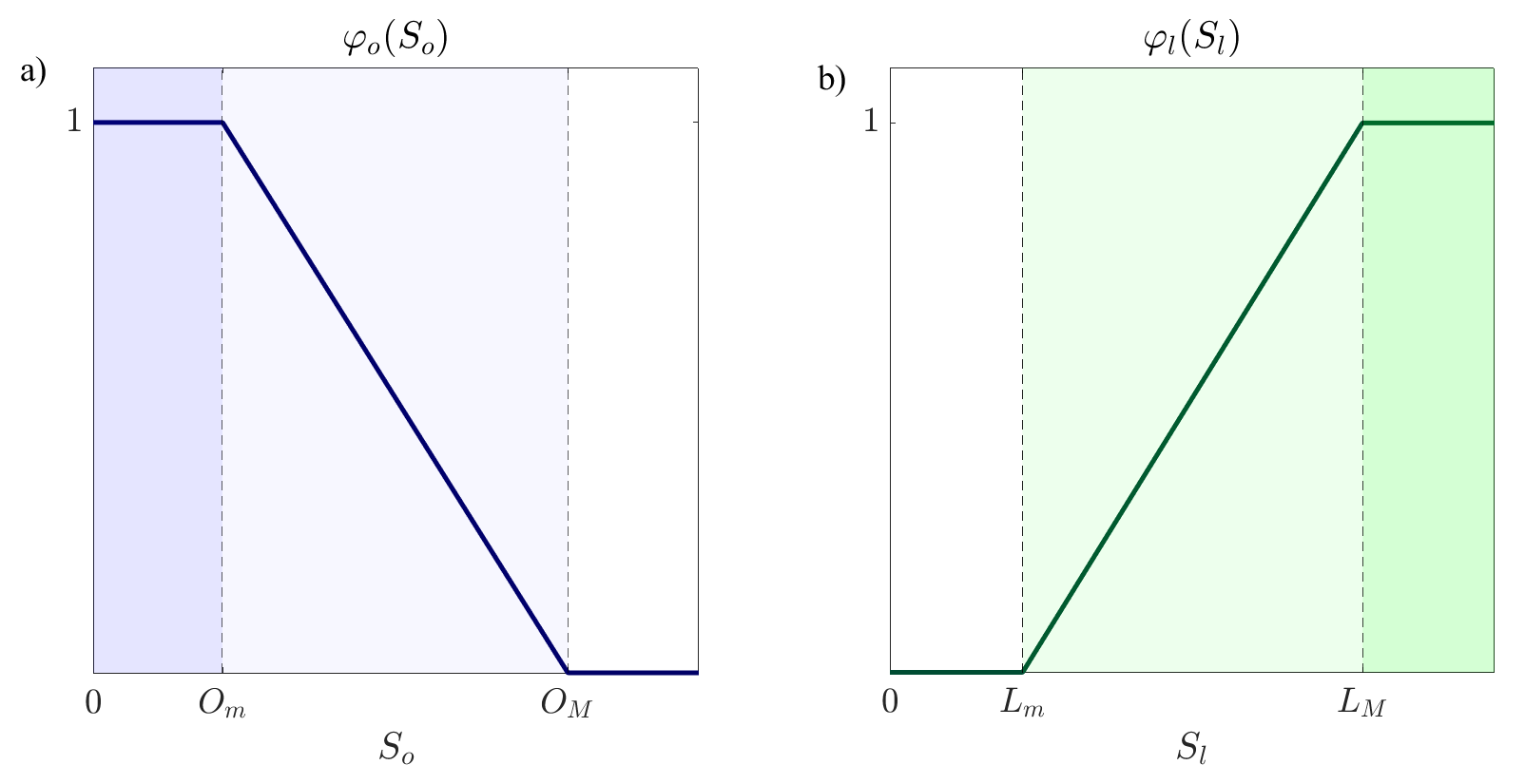}
	\caption{\footnotesize{{\bf a)} Plot of the fittest level of expression of the hypoxia-resistant gene $\varphi_o(S_o)$ defined via~\eqref{phio_So}. The vertical, dashed lines highlight the oxygen levels $O_m$ and $O_M$. Hence, the white region corresponds to normoxic conditions, the pale-blue region corresponds to moderately-oxygenated environments and the blue region corresponds to hypoxic conditions. {\bf b)} Plot of the fittest level of expression of the acidity-resistant gene $\varphi_l(S_l)$ defined via~\eqref{phil_Sl}. The vertical, dashed lines highlight the lactate levels $L_m$ and $L_M$. Hence, the white region corresponds to mildly-acidic conditions, the pale-green region corresponds to moderately-acidic conditions and the green region corresponds to highly-acidic conditions.}}
	\label{figure2}
\end{figure}

Under Assumptions~\ref{Ass4} and~\ref{Ass5}, we define the rate of cell death due to lactate-driven selection $D(S_l,y_1)$ as
\begin{equation}
D(S_l, y_1) := \eta_l \, \big(y_1-{\varphi_l(S_l)}\big)^2.
\label{death}
\end{equation}
In~\eqref{death}, the parameter $\eta_l>0$ is a selection gradient that quantifies the intensity of lactate-driven selection and the function $\varphi_l(S_l)$ is the fittest level of expression of the acidity-resistant gene under the environmental conditions given by the lactate concentration $S_l$. Considerations analogous to those made in Remark~\ref{remark} on the term $\big(y_2-\varphi_o(S_o)\big)^2$ in~\eqref{proliferation} apply to the term $\big(y_1-\varphi_l(S_l)\big)^2$ in~\eqref{death}. Finally, we use the definition of the function $\varphi_l(S_l)$ given hereafter ({\it cf.} the plot in Figure~\ref{figure2}b), so that Assumption~\ref{Ass6} is satisfied:
\begin{equation}
\varphi_l (S_l):=\begin{cases}
\displaystyle
0 \qquad  \qquad \quad  \quad S_l \leq L_m\\
\\
\displaystyle
\frac{S_l-L_m}{L_M-L_m} \quad  \quad L_m < S_l < L_M\\
\\
\displaystyle
1 \qquad  \qquad \quad  \quad S_l \geq L_M.
\end{cases}
\label{phil_Sl}
\end{equation}

\subsubsection{Modelling competition for space}
We define the rate of cell death due to competition for space associated with saturation of the cell density as
\begin{equation}	
d(\rho) := \kappa \, \rho.	
\end{equation}	
Here, the proportionality constant $\kappa>0$ is related to the local carrying capacity of the tumour, which may vary depending on the tumour type.

\subsection{Dynamics of abiotic factors}
Oxygen and glucose are consumed by tumour cells, while lactate is produced by tumour cells as a waste product of glycolysis. Moreover, oxygen, glucose and lactate diffuse in space and decay over time. Hence, their dynamics are governed by the following balance equations for the functions $S_o(t,x)$, $S_g(t,x)$ and $S_l(t,x)$, respectively, 
\begin{equation}
\frac{\partial S_o}{\partial t} \; = \; \underbrace{\beta_o\frac{\partial^2 S_o}{\partial x^2}}_{\substack{\text{diffusion} }} \quad - \; \underbrace{\lambda_o\,S_o}_{\substack{\text{natural} \\\text{decay}}}\quad - \; \,\underbrace{\zeta_o \,\, p_o(S_o) \,\,\rho,}_{\text{consumption by tumour cells}} 
\label{oxygen}
\end{equation}
\begin{equation}
\frac{\partial S_g}{\partial t}\; = \;\underbrace{\beta_g\frac{\partial^2 S_g}{\partial x^2}}_{\substack{\text{diffusion} }} \quad - \;\underbrace{\lambda_g\,S_g}_{\substack{\text{natural} \\\text{decay}}}\quad - \;\underbrace{\zeta_g\,\, p_g(S_o,S_g) \,\,\rho}_{\text{consumption by tumour cells}}
\label{glucose}
\end{equation}
and
\begin{equation}
\frac{\partial S_l}{\partial t}\; = \;\underbrace{\beta_l\frac{\partial^2 S_l}{\partial x^2}}_{\substack{\text{diffusion} }} \quad - \; \underbrace{\lambda_l\,S_l}_{\substack{\text{natural} \\\text{decay}}}\quad + \;\,\underbrace{\zeta_l \,\, p_g(S_o,S_g)\,\,\rho,}_{\text{production by tumour cells}}
\label{lactate}
\end{equation}
with $(t,x) \in (0,{\rm T}] \times (0,{\rm L})$, subject to suitable boundary conditions (see considerations below) and initial conditions. 

In~\eqref{oxygen}-\eqref{lactate}, the parameters $\beta_o>0$, $\beta_g>0$ and $\beta_l>0$ are the diffusion coefficients of oxygen, glucose and lactate, respectively, while the parameters $\lambda_o>0$, $\lambda_g>0$ and $\lambda_l>0$ are the natural decay rates of the three abiotic factors. The third term on the right-hand side of~\eqref{oxygen} is the consumption rate of oxygen by tumour cells, which is proportional to the product between the cell density $\rho$ and the rate of cell proliferation via oxidative phosphorylation $p_o(S_o)$, which is defined via~\eqref{def:popg}. The parameter $\zeta_o>0$ is a conversion factor linked to oxygen consumption by the cells. Analogous considerations hold for the third term on the right-hand side of~\eqref{glucose}, which models the consumption rate of glucose by tumour cells. Furthermore, the third term on the right-hand side of~\eqref{lactate} is the production rate of lactate by tumour cells, which is assumed to be proportional to the product between the cell density $\rho$ and the rate of cell proliferation via glycolysis $p_g(S_o,S_g)$ defined via~\eqref{def:popg}. The constant of proportionality is the conversion factor $\zeta_l>0$ linked to lactate production by the cells. 

We assume that oxygen and glucose enter the spatial domain through the blood vessel only, while lactate is flushed out through the blood vessel only. Hence, focussing on the case where the inflow rate of oxygen and glucose and the outflow rate of lactate are constant, we complement~\eqref{oxygen}-\eqref{lactate} with the following boundary conditions at $x=0$
\begin{equation}
S_o(t,0) = \overline{S}_o, \quad S_g(t,0) = \overline{S}_g, \quad S_l(t,0) = \underline{S}_l \quad \text{for all } t >0,
\label{eq:BCs0}
\end{equation}
where $\overline{S}_o>0$ and $\overline{S}_g>0$ are related to the average physiological levels of oxygen and glucose in proximity to blood vessels, while $\underline{S}_l>0$ is a small  parameter of value close to zero. In particular, we have $\overline{S}_o>O_M$ and $\underline{S}_l < L_m$. Moreover, we assume that far from the blood vessel the concentrations of oxygen and glucose drop to some lower values $0<\underline{S}_o<\overline{S}_o$ and $0<\underline{S}_g<\overline{S}_g$, which correspond to the levels of oxygen and glucose which are typically observed in regions distant from blood vessels. In particular, we have $\underline{S}_o<O_m$. Furthermore, we model abnormal accumulation of lactate, which is expected to occur far from blood vessels, imposing zero-flux boundary conditions. Therefore, we complement~\eqref{oxygen}-\eqref{lactate} with the following boundary conditions at $x={\rm L}$
\begin{equation}
S_o(t,{\rm L}) = \underline{S}_o, \quad S_g(t,{\rm L}) = \underline{S}_g, \quad \dfrac{\partial S_l(t,{\rm L})}{\partial x} = 0 \quad \text{for all } t >0.
\label{eq:BCsX}
\end{equation}    

\section{Main results}
\label{Sec3}
In this section, we present the results of numerical simulations of the mathematical model defined by~\eqref{main} coupled with \eqref{oxygen}-\eqref{lactate} and we discuss their biological relevance. First, we describe the set-up of numerical simulations (see Section~\ref{Sec:setup}). Next, we present a sample of numerical solutions that summarise the spatial dynamics of tumour cells and abiotic factors (see Section~\ref{tumour growth}). Then, we report on the results of numerical simulations showing the evolutionary dynamics of tumour cells and the emergence of phenotypic heterogeneity (see Section~\ref{adaptive dynamics}). Finally, we present the results of numerical simulations that reveal the existence of alternative evolutionary pathways that may lead to the development of resistance to hypoxia and acidity in vascularised tumours (see Section~\ref{evolutionary}).

\subsection{Set-up of numerical simulations}	
\label{Sec:setup}
Numerical simulations are carried out assuming ${\rm L}=400 \,\mu m $, which is chosen coherently with experimental data reported in~\cite{Molavian9141}, and $t \in [0,{\rm T}]$, where the final time ${\rm T}>0$ is such that the solutions of the model equations are at numerical equilibrium for $t={\rm T}$.

\begin{figure}[htp!]
	\centering
	\includegraphics[width=0.5\textwidth]{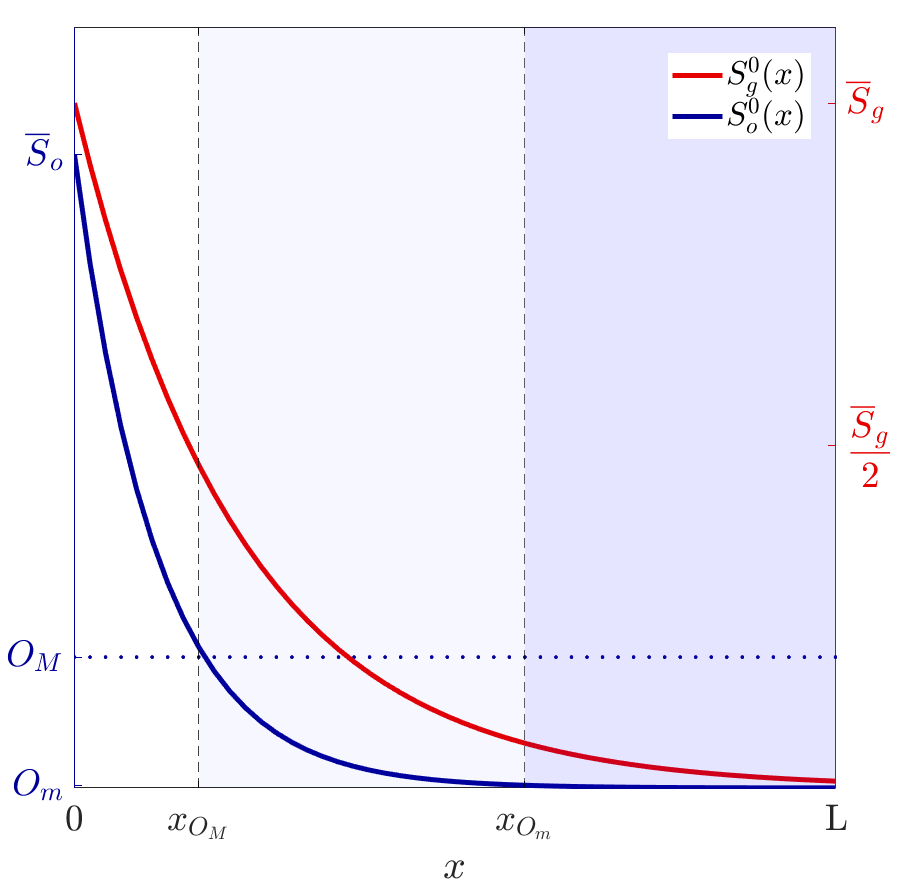}
	\caption{\footnotesize{Plots of the initial distribution of oxygen $S^0_o(x)$ (blue curve, axis on the left) and the initial distribution of glucose $S^0_g(x)$ (red curve, axis on the right) which are defined in such a way as to match the experimental equilibrium distributions of oxygen and glucose presented in~\cite[Fig. 2]{Molavian9141}. The vertical, dashed lines highlight the points $x_{O_M}$ and $x_{O_m}$ such that $S^0_o(x_{O_M})=O_M$ and $S^0_o(x_{O_m})=O_m$. Hence, the white region corresponds to normoxic conditions, the pale-blue region corresponds to moderately-oxygenated environments and the blue region corresponds to hypoxic conditions.}}
	\label{figure3}
\end{figure}

\paragraph{Initial conditions.} We consider~\eqref{oxygen}, \eqref{glucose} and \eqref{lactate} subject, respectively, to the following initial conditions
\begin{equation}
\label{icS}
S_o(0,x) = S^0_o(x), \quad  S_g(0,x) = S^0_g(x) \quad \text{and} \quad S_l(0,x) =  S^0_l(x) \equiv \underline{S}_l.
\end{equation}   
Here, the functions $S^0_o(x)$ and $S^0_g(x)$ (see Figure~\ref{figure3}) are defined in such a way as to match the experimental equilibrium distributions of oxygen and glucose presented in~\cite[Fig. 2]{Molavian9141}, while $\underline{S}_l$ is the same small parameter used in~\eqref{eq:BCs0}, \emph{i.e.} $\underline{S}_l<L_m$. Initial conditions~\eqref{icS} correspond to a situation in which the initial distributions of oxygen and glucose match with experimental equilibrium distributions of such abiotic factors and lactate is present at a uniform level which is below the threshold level $L_m$, that is, the level below which the environment surrounding the cells is mildly acidic and the fittest level of expression of the acidity-resistant gene is the minimal one. Moreover, we complement~\eqref{main} with the following initial condition
\begin{equation}
\label{icn}
n(0,x,{\bf y}) = n^0(x,{\bf y}) := 200 \ \exp\bigg({-\frac{(x-0)^2}{0.0002}-\frac{|{\bf y} - 0|^2}{0.4}}\bigg),
\end{equation}   
which corresponds to a biological scenario in which at the initial time $t=0$ most tumour cells are concentrated near the blood vessel and are characterised by the minimal expression level of both the hypoxia-resistant gene and the acidity-resistant gene.  

\paragraph{Boundary conditions.} 
We use the following values of the parameters $\overline{S}_o$, $\overline{S}_g$ and $\underline{S}_l$ in~\eqref{eq:BCs0}
\begin{equation}
\overline{S}_o = 2.08 \times 10^{-6} \, g/cm^3, \quad \overline{S}_g=1.35 \times 10^{-4} \, g/cm^3, \quad \underline{S}_l= 10^{-8} \, g/cm^3
\label{eq:BCs0num}
\end{equation}   
and the following values of the parameters $\underline{S}_o$ and $\underline{S}_g$ in~\eqref{eq:BCsX}
\begin{equation}
\underline{S}_o = 2 \times 10^{-10}\, g/cm^3, \quad \underline{S}_g= 1.35 \times 10^{-6} \, g/cm^3.
\label{eq:BCsXnum}
\end{equation}   
The values of $\overline{S}_o$ and $\overline{S}_g$ correspond to the average physiological levels of oxygen and glucose in proximity to blood vessels reported in~\cite{Molavian9141}. Moreover, the values of $\underline{S}_o$ and $\underline{S}_g$ correspond to the $0.1\%$ of $\overline{S}_o$ and the $1\%$ of $\overline{S}_g$, respectively. This is because, based on experimental data reported in~\cite{Molavian9141}, we expect the concentrations of oxygen and glucose at $400\,\mu m$ from the blood vessel (\emph{i.e.} at $x={\rm L}$) to drop, respectively, below the $0.1\%$ and the $1\%$ of their value near the blood vessel.
\paragraph{Parameter values.} 
Unless otherwise explicitly stated, we use the values of the model parameters listed in Table~\ref{tableparameters}, which are chosen to be consistent with the existing literature, except for the values of the parameters $\eta_o$, $\eta_l$, $\lambda_l$ and $\zeta_l$ that are model specific in that we could not find them in the literature and are defined on the basis of the following considerations. The value of the conversion factor for lactate production $\zeta_l$ is chosen to be the same as the value of the conversion factor for glucose consumption $\zeta_g$. Furthermore, the value of the rate of natural decay of lactate $\lambda_l$ is such that the distribution of lactate at numerical equilibrium (\emph{i.e.} the graph of $S_l({\rm T},x)$) is similar to the lactate distributions reported in~\cite{Molavian9141}. \textcolor{black}{Finally, although the values of the selection gradients $\eta_o$ and $\eta_l$ are chosen with exploratory aim, a systematic sensitivity analysis of the evolutionary dynamics of tumour cells to the values of these parameters was carried out, and the key findings from such sensitivity analysis are summarised by the results presented in Section~\ref{evolutionary}.} \textcolor{black}{We also note that the value of the rate of phenotypic changes given in Table~\ref{tableparameters} is consistent with experimental data reported in~\cite{doerfler2006dna} and~\cite{duesberg2000explaining}.}
\begin{table}[htp!]
\footnotesize{
				\caption{\footnotesize{Parameter values used in numerical simulations.}}
		\begin{tabular}{l l l l c}
		\hline\noalign{\smallskip}
			Par.	& Biological meaning  & Value \qquad  &  Units  & Ref.\\
		\noalign{\smallskip}\hline\noalign{\smallskip}
			\, $\beta_o$  & Diffusion coefficient of oxygen & $1.46\,\times\,10^{-5}$& $cm^2 s^{-1}$ & \cite{Molavian9141} \\
			\, $\beta_g$  & Diffusion coefficient of glucose & $1.10\,\times\,10^{-6}$ & $cm^2 s^{-1}$ & \cite{Molavian9141} \\
			\, $\beta_l$  & Diffusion coefficient of lactate &$1.9\,\times\,10^{-6}$& $cm^2 s^{-1}$ & \cite{Molavian9141} \\
			\, $\beta_n$  & Cell motility & $10^{-13}$ & $cm^2 s^{-1}$ & \cite{villa2021modeling} \\
			\, $\theta$ & Rate of phenotypic changes & $10^{-13}$ & $s^{-1}$& \cite{villa2021modeling} \\
			\, $\lambda_o$  & Rate of natural decay of oxygen & $1.2\,\times\,10^{-3}$ & $s^{-1}$ & \cite{Molavian9141} \\
			\, $\lambda_g$  & Rate of natural decay of glucose & $2.33\,\times\,10^{-5}$ & $s^{-1}$& \cite{Molavian9141} \\
			\, $\lambda_l$  & Rate of natural decay of lactate & $5\,\times\,10^{-2}$ & $s^{-1}$& {\it ad hoc} \\
			\, $\gamma_o$  & Max. prolif. rate via oxidative phosphorylation & $3.65\,\times\,10^{-7}$ & $s^{-1}$ & \cite{Molavian9141} \\
			\, $\alpha_o$  & Michaelis-Menten constant of oxygen & $6.4\,\times\,10^{-9}$ & $g \ cm^{-3}$ & \cite{Molavian9141} \\
			\, $\gamma_g$  & Max. prolif. rate via glycolysis & $3.42\,\times\,10^{-7}$ &$s^{-1}$ & \cite{Molavian9141} \\
			\, $\alpha_g$  & Michaelis-Menten constant of glucose & $9\,\times\,10^{-6}$ & $g \ cm^{-3}$& \cite{Molavian9141} \\
			\, $\eta_o$  & Selection gradient related to oxygen & $3.65\,\times\,10^{-2}$ & $s^{-1}$  & {\it ad hoc}\\
			\, $\eta_l$  & Selection gradient related to lactate &$10^{-2}$ & $s^{-1}$ & {\it ad hoc} \\
			\, $\kappa$  & Rate of cell death due to competition for space & $2\,\times\,10^{-10}$ & $cm^{3} s^{-1} cells^{-1}$ & \cite{LORENZI2018101}\\
			\, $\zeta_o$  & Conversion factor for oxygen consumption &$10^{-8}$& $g \ cells^{-1}$ & \cite{LORENZI2018101} \\
			\, $\zeta_g$  & Conversion factor for glucose consumption &$10^{-8}$ & $g \ cells^{-1}$ & \cite{LORENZI2018101}\\
			\, $\zeta_l$  & Conversion factor for lactate production &$10^{-8}$ & $g \ cells^{-1}$ & {\it ad hoc} \\
			\, $O_m$  & Threshold level of oxygen for hypoxic env. & $8.2 \,\times\, 10^{-9}$ & $g \ cm^{-3}$ & \cite{Treatment} \\
			\, $O_M$  & Threshold level of oxygen for normoxic env. & $4.3 \,\times\, 10^{-7}$ & $g \ cm^{-3}$ & \cite{Treatment} \\
			\, $L_m$  & Threshold level of lactate for mildly-acidic env. & $2 \,\times\, 10^{-5}$ & $g \ cm^{-3}$ & \cite{Robertson-Tessi1567} \\	     
			\, $L_M$  &  Threshold level of lactate for highly-acidic env. & $7.15 \,\times\, 10^{-5}$ & $g \ cm^{-3}$ & \cite{Robertson-Tessi1567} \\
			\, ${\rm L}$  &  Max. distance from blood vessel & $400$ & $\mu m$ & \cite{Molavian9141} 	\\
		\noalign{\smallskip}\hline
		\end{tabular}
		\label{tableparameters}
}
\end{table}
\paragraph{Numerical methods.} 
Numerical solutions are constructed using a uniform discretisation of the interval $[0,{\rm L}]$ as the computational domain of the independent variable $x$ and a uniform discretisation of the square $[0,1]^2$ as the computational domain of the independent variable ${\bf y}$. We also discretise the interval $[0,{\rm T}]$ with a uniform step. The method for constructing numerical solutions  is based on an explicit finite difference scheme in which a three-point and a five-point stencils are used to approximate the diffusion terms in $x$ and ${\bf y}$, respectively, and an implicit-explicit finite difference scheme is used for the reaction terms~\cite{leveque2007finite,lorz2013populational}. All numerical computations are performed in {\sc Matlab}.

\subsection{Dynamics of the cell density and the concentrations of abiotic factors}
\label{tumour growth}
\begin{figure}[htp!]
	\centering
	\includegraphics[width=1.02\textwidth]{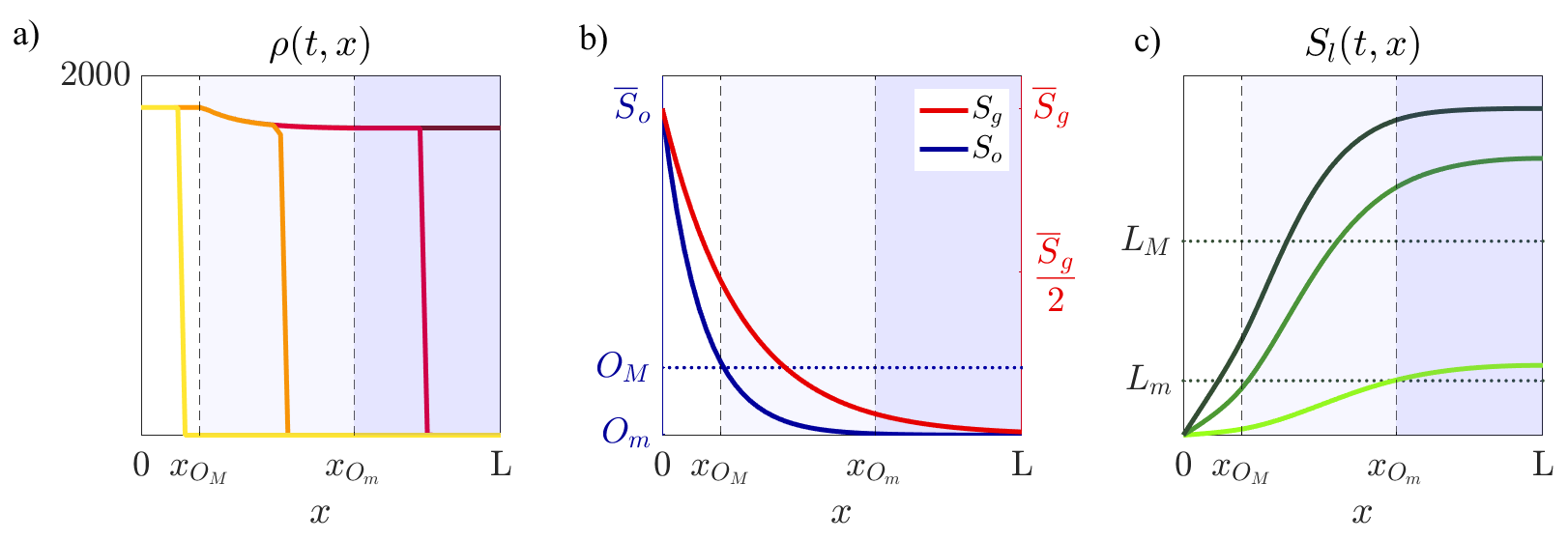}
	\caption{\footnotesize{{\bf a)} Plots of the cell density $\rho(t,x)$ at four successive time instants (yellow, orange, red and  burgundy lines). The burgundy line highlights $\rho({\rm T},x)$. {\bf b)} Plots of  the concentrations of oxygen $S_o({\rm T},x)$ (blue line) and glucose $S_g({\rm T},x)$ (red line). {\bf c)} Plots of the concentration of lactate $S_l(t,x)$ at three successive time instants (light-green, green and dark-green lines). The dark-green line highlights $S_l({\rm T},x)$. In every panel, the vertical, dashed lines highlight the points $x_{O_M}$ and $x_{O_m}$ such that $S_o({\rm T},x_{O_M})=O_M$ and $S_o({\rm T},x_{O_m})=O_m$. Hence, the white region corresponds to normoxic conditions, the pale-blue region corresponds to moderately-oxygenated environments and the blue region corresponds to hypoxic conditions.}}
	\label{figure4}
\end{figure}

The dynamics of the density of tumour cells and the concentrations of abiotic factors are illustrated by the plots in Figure~\ref{figure4}. In summary, the cell density and the concentration of lactate at successive times (\emph{i.e.} the graphs of $\rho(t,x)$ for four different values of $t$ and $S_l(t,x)$ for three different values of $t$) are displayed in Figure~\ref{figure4}a and Figure~\ref{figure4}c, respectively, while the concentrations of oxygen and glucose at time ${\rm T}$ (\emph{i.e.} the graphs of $S_o({\rm T},x)$ and $S_g({\rm T},x)$) are displayed in Figure~\ref{figure4}b. 

The dashed lines in Figure~\ref{figure4} highlight the spatial positions $x_{O_M}$ and $x_{O_m}$ at which the oxygen concentration at time ${\rm T}$ crosses, respectively, the threshold values $O_M$ and $O_m$ (\emph{i.e.} $S_o({\rm T},x_{O_M})=O_M$ and $S_o({\rm T},x_{O_m})=O_m$). Hence, the white region (\emph{i.e.} the interval $[0,x_{O_M}]$), the pale-blue region (\emph{i.e.} the interval $(x_{O_M},x_{O_m})$) and the blue region (\emph{i.e.} the interval $[x_{O_m}, {\rm L}]$) correspond to normoxic, moderately-oxygenated and hypoxic environmental conditions, respectively.

\begin{figure}[htp!]
	\centering
	\includegraphics[width=0.7\textwidth]{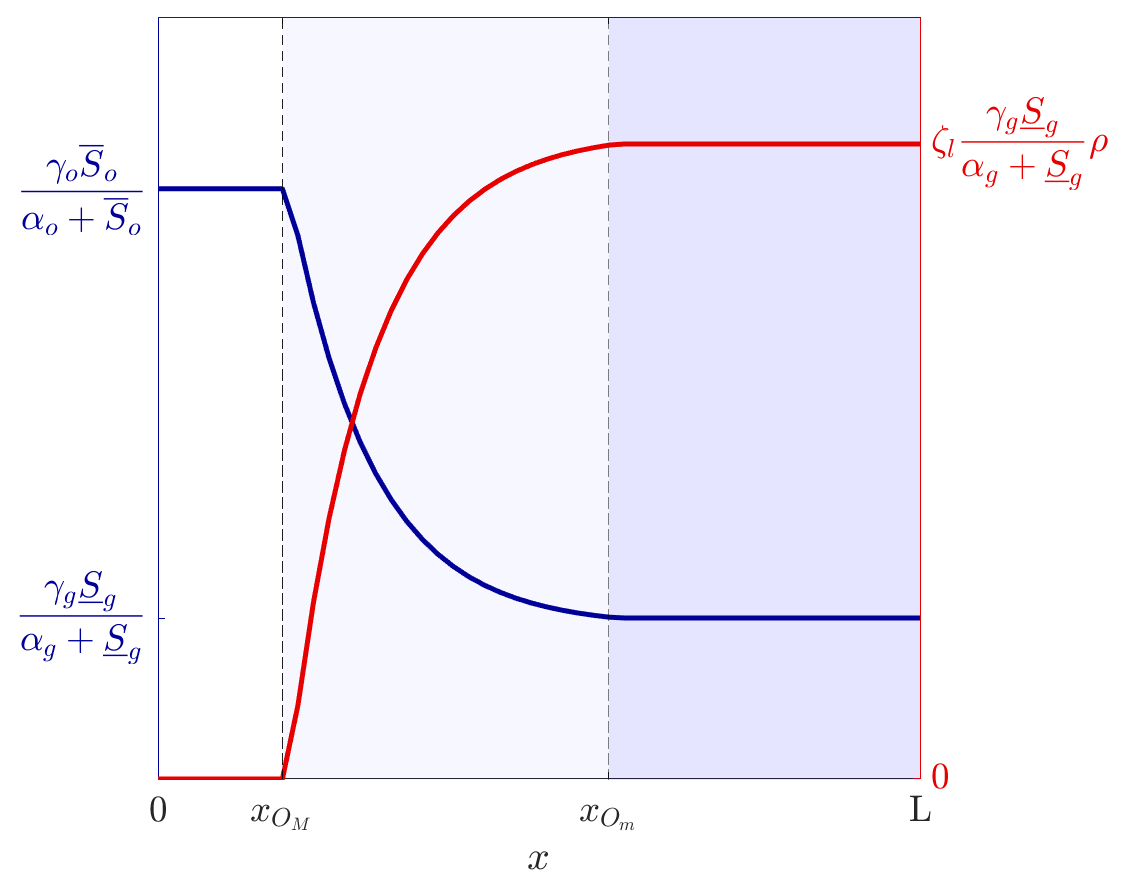}
	\caption{\footnotesize{Plots of the cell proliferation rate $p_o(S_o({\rm T},x))+p_g(S_o({\rm T},x),S_g({\rm T},x))$ (blue curve, axis on the left) and the production rate of lactate $\zeta_l \, p_g(S_o({\rm T},x),S_g({\rm T},x))\,\rho({\rm T},x)$ (red curve, axis on the right) defined via~\eqref{def:popg}. The vertical, dashed lines highlight the points $x_{O_M}$ and $x_{O_m}$ such that $S_o({\rm T},x_{O_M})=O_M$ and $S_o({\rm T},x_{O_m})=O_m$. Hence, the white region corresponds to normoxic conditions, the pale-blue region corresponds to moderately-oxygenated environments and the blue region corresponds to hypoxic conditions.}}
	\label{figure5}
\end{figure}

\textcolor{black}{The curves in Figure~\ref{figure4}a summarise the evolution of the cell number density $\rho(t,x)$, which behaves like an invading front whereby growth is saturated at a value that decreases with the distance from the blood vessel (\emph{i.e.} $\rho(t,x)$ converges to a form of generalised transition wave~\cite{berestycki2012generalized,berestycki2015asymptotic}). These results illustrate how the synergistic interaction between cell proliferation, which occurs until the local carrying capacity of the tissue is reached, and cell movement allows tumour cells, which are initially located in the proximity of the blood vessel, to invade the surrounding tissue.} \textcolor{black}{The result that the plateau value of the cell number density decreases with the distance from the blood vessel, which is a result with broad structural stability under parameter changes (see Appendix~\ref{sec:ApA}), reflects the fact that, in our model, the cell proliferation rate in normoxic conditions, whereby the energy needed for cell proliferation is produced via oxidative phosphorylation, is higher than the cell proliferation rate in moderately-oxygenated environments, which in turn is higher than the cell proliferation rate in hypoxic conditions, whereby the energy needed for cell proliferation is produced via glycolysis ({\it cf.} the blue curve in Figure~\ref{figure5}). This is in line with experimental evidence indicating that glycolysis is less efficient than oxidative phosphorylation as a mechanism to produce the energy required for cell proliferation.}

Moreover, the curves in Figure~\ref{figure4}b show how the reaction-diffusion dynamics of oxygen and glucose, along with the inflow through the blood vessel and the consumption by tumour cells, lead the concentrations of such abiotic factors to converge to some stable values which decrease as the distance from the blood vessel increases (\emph{i.e.} at $t={\rm T}$, $S_o(t,x)$ and $S_g(t,x)$ appear to be at numerical equilibrium and are monotonically decreasing functions of $x$). Notice that the distributions of oxygen and glucose at the final time ${\rm T}$ are close to the initial distributions $S^0_o(x)$ and $S^0_g(x)$ displayed in Figure~\ref{figure3}. This is to be expected. In fact, since $S^0_o(x)$ and $S^0_g(x)$ are defined in such a way as to match experimental equilibrium distributions of oxygen and glucose, under the biologically informed parameter values ({\it cf.} Table~\ref{tableparameters}) and boundary conditions ({\it cf.} \eqref{eq:BCs0}, \eqref{eq:BCsX} and \eqref{eq:BCs0num}, \eqref{eq:BCsXnum}) used here, the concentrations $S_o(t,x)$ and $S_g(t,x)$ reach quickly numerical equilibrium. We verified via additional numerical simulations (results not shown) that, as one would expect, the concentrations of oxygen and glucose at numerical equilibrium do not depend on the choice of the initial conditions.

\textcolor{black}{Finally, the curves in Figure~\ref{figure4}c summarise the evolution of the concentration of lactate $S_l(t,x)$, which is the result of the interplay between the production of this abiotic factor by tumour cells, its reaction-diffusion dynamics and its outflow through the blood vessel. It is known that, as a waste product of glycolysis, lactate is mainly produced and accumulate in moderately-oxygenated and hypoxic regions, where cell proliferation relies more on glycolysis and tissue perfusion is poorer. In agreement with this, the curves in Figure~\ref{figure4}c demonstrate that the concentration of lactate increases with the distance from the blood vessel. In particular, the values attained by $S_l({\rm T},x)$, which depend on the values of the production rate of lactate in our model ({\it cf.} the red curve in Figure~\ref{figure5}), are in agreement with lactate concentrations reported in~\cite{Molavian9141}.}

\subsection{Evolutionary dynamics of tumour cells and emergence of phenotypic heterogeneity}
\label{adaptive dynamics}
As discussed in the previous section, reaction-diffusion dynamics of abiotic factors and mutual interactions between abiotic factors and tumour cells lead to the emergence of spatial variations in the concentrations of oxygen and lactate -- {\it i.e.} the oxygen concentration $S_o({\rm T},x)$ is a monotonically decreasing function of $x$ while the lactate concentration $S_l({\rm T},x)$ is a monotonically increasing function of $x$ ({\it cf.} plots in Figure~\ref{figure4}b and Figure~\ref{figure4}c).

\begin{figure}[htp!]
	\centering
	\includegraphics[width=0.9\textwidth]{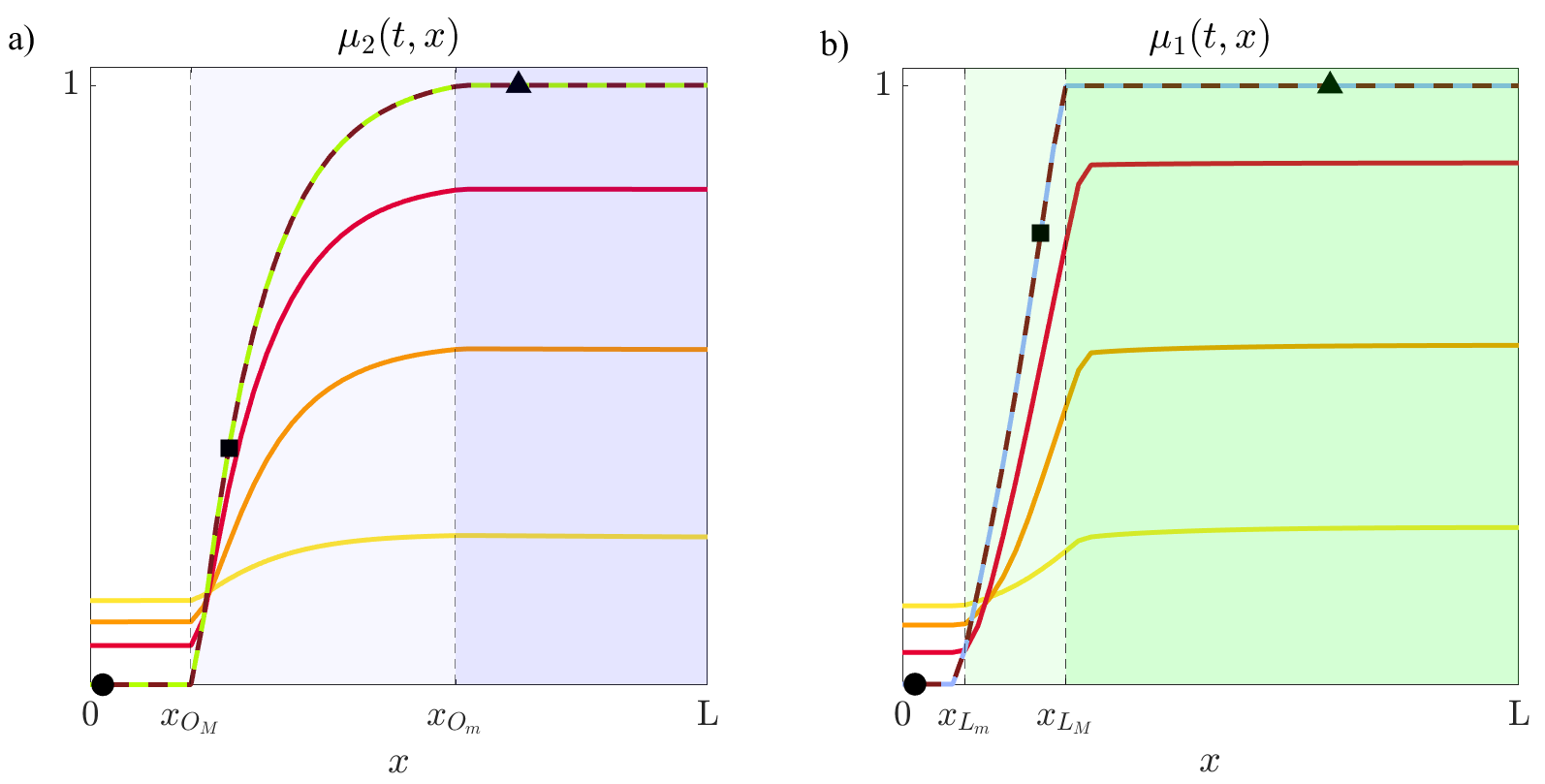}
	\includegraphics[width=1\textwidth]{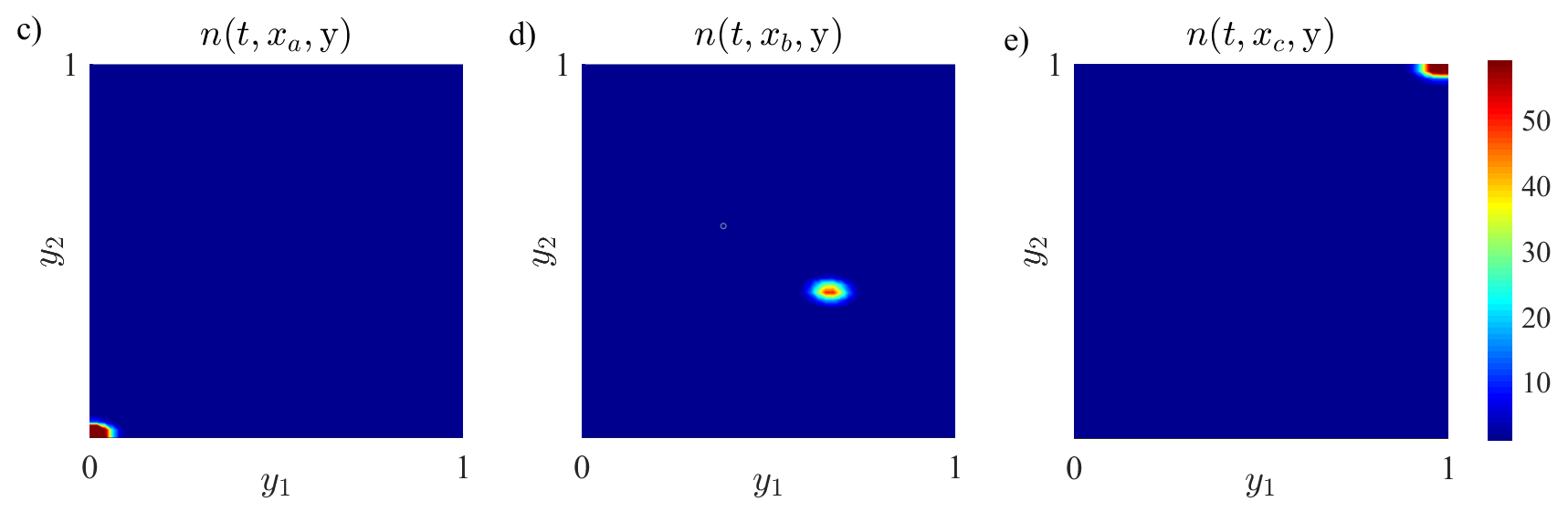}
	\caption{\footnotesize{{\bf a)} Plots of the normalised level of expression of the hypoxia-resistant gene $\mu_2(t,x)$  at four successive time instants (yellow, orange, red and light-green lines). The light-green line highlights $\mu_2({\rm T},x)$ and the burgundy, dashed line highlights the fittest level of expression of the hypoxia-resistant gene $\varphi_o(S_o({\rm T},x))$ defined via~\eqref{phio_So}. The vertical, dashed lines highlight the points $x_{O_M}$ and $x_{O_m}$ such that $S_o({\rm T},x_{O_M})=O_M$ and $S_o({\rm T},x_{O_m})=O_m$. Hence, the white region corresponds to normoxic conditions, the pale-blue region corresponds to moderately-oxygenated environments and the blue region corresponds to hypoxic conditions. {\bf b)} Plots of the normalised level of expression of the acidity-resistant gene $\mu_1(t,x)$ at four successive time instants (yellow, orange, red and light-blue lines). The light-blue line highlights $\mu_1({\rm T},x)$ and the burgundy, dashed line highlights the fittest level of expression of the acidity-resistant gene $\varphi_l(S_l({\rm T},x))$ defined via~\eqref{phil_Sl}. The vertical, dashed lines highlight the points $x_{L_m}$ and $x_{L_M}$ such that $S_l({\rm T},x_{L_m})=L_m$ and $S_l({\rm T},x_{L_M})=L_M$. Hence, the white region corresponds to mildly-acidic conditions, the pale-green region corresponds to moderately-acidic conditions and the green region corresponds to highly-acidic conditions. {\bf c)} - {\bf e)} Plots of the local phenotypic distribution of tumour cells $n({\rm T},x,{\bf y})$ at the points $x=x_a$, $x=x_b$ and $x=x_c$ highlighted, respectively, by the circle, square and triangle markers shown in panels a) and b).}}
	\label{figure6}
\end{figure}

Spatial variability of oxygen and lactate concentrations can lead to the formation of environmental gradients resulting in the selection for cells with phenotypic characteristics that vary with distance from the blood vessel, thus promoting the emergence of intra-tumour phenotypic heterogeneity. \textcolor{black}{The numerical results displayed in Figure~\ref{figure6} support the idea that this can be effectively captured by our model.}

The dashed lines in Figure~\ref{figure6}a and Figure~\ref{figure6}b highlight the fittest levels of expression of the hypoxia-resistant gene (see Figure~\ref{figure6}a) and the acidity-resistant gene (see Figure~\ref{figure6}b) at time ${\rm T}$ [\emph{i.e.} the graphs of $\varphi_o(S_o({\rm T},x))$ and $\varphi_l(S_l({\rm T},x))$], while the solid lines display the local mean levels of expression of the hypoxia-resistant gene (see Figure~\ref{figure6}a) and the acidity-resistant gene (see Figure~\ref{figure6}b) at four successive times (\emph{i.e.} the graphs of $\mu_2(t,x)$ and $\mu_1(t,x)$ for four different values of $t$). In analogy with Figure~\ref{figure4}, the vertical, dashed lines in Figure~\ref{figure6}a highlight the spatial positions $x_{O_M}$ and $x_{O_m}$ at which the oxygen concentration at time ${\rm T}$ crosses, respectively, the threshold values $O_M$ and $O_m$ (\emph{i.e.} $S_o({\rm T},x_{O_M})=O_M$ and $S_o({\rm T},x_{O_m})=O_m$). Hence, the white region corresponds to normoxic conditions, the pale-blue region corresponds to moderately-oxygenated environments and the blue region corresponds to hypoxic conditions. Moreover, the vertical, dashed lines in Figure~\ref{figure6}b highlight the spatial positions $x_{L_m}$ and $x_{L_M}$ at which the lactate concentration at time ${\rm T}$ crosses, respectively, the threshold values $L_m$ and $L_M$ (\emph{i.e.} $S_l({\rm T},x_{L_m})=L_m$ and $S_l({\rm T},x_{L_M})=L_M$). Hence, the white region corresponds to mildly-acidic conditions, the pale-green region corresponds to moderately-acidic conditions and the green region corresponds to highly-acidic conditions. 

As shown by the plots in Figure~\ref{figure6}a and Figure~\ref{figure6}b, the local mean levels of expression of the hypoxia- and acidity-resistant genes converge, as time passes, to the fittest ones (\emph{i.e.} $\mu_2({\rm T},x)$ matches with $\varphi_o(S_o({\rm T},x))$ and $\mu_1({\rm T},x)$ matches with $\varphi_l(S_l({\rm T},x)$), the values of which vary with the distance from the blood vessel depending on the local concentrations of oxygen and lactate. In more detail, the local mean level of expression of the hypoxia-resistant gene at time ${\rm T}$ is the minimal one in normoxic conditions (\emph{i.e.} $\mu_2({\rm T},x) \equiv 0$ for $x \in [0,x_{O_M}]$), the maximal one in hypoxic conditions (\emph{i.e.} $\mu_2({\rm T},x) \equiv 1$ for $x \in [x_{O_m},{\rm L}])$ and increases with the oxygen concentration in moderately-oxygenated environments (\emph{i.e.} $\mu_2({\rm T},x)$ increases monotonically from $0$ to $1$ for $x \in (x_{O_M},x_{O_m})$). Furthermore, the local mean level of expression of the acidity-resistant gene at time ${\rm T}$ is the minimal one in the mildly-acidic region (\emph{i.e.} $\mu_1({\rm T},x) \equiv 0$ for $x \in [0,x_{L_m}]$), the maximal one in highly-acidic conditions (\emph{i.e.} $\mu_1({\rm T},x) \equiv 1$ for $x \in [x_{L_M},{\rm L}])$ and increases with the lactate concentration in moderately-acidic environments (\emph{i.e.} $\mu_1({\rm T},x)$ increases monotonically from $0$ to $1$ for $x \in (x_{L_m},x_{L_M})$). \textcolor{black}{These are results with broad structural stability under parameter changes (see Appendix~\ref{sec:ApA}).}

Finally, the plots in Figures~\ref{figure6}c-e show that, at every position $x \in [0,{\rm L}]$, the local phenotypic distribution of tumour cells at time ${\rm T}$ (\emph{i.e.} the local population density function $n({\rm T},x,{\bf y})$) is unimodal and attains its maximum at the fittest phenotypic state ${\bf y} = \big(\varphi_o(S_o({\rm T},x)),\varphi_l(S_l({\rm T},x) \big)$. \textcolor{black}{As discussed in Appendix~\ref{sec:ApA}, such a qualitative behaviour of $n({\rm T},x,{\bf y})$ is in agreement with predictions based on the formal asymptotic results presented in~\cite{villa2021modeling}.}

\begin{figure}[htp!]
	\centering
	\includegraphics[width=1\textwidth]{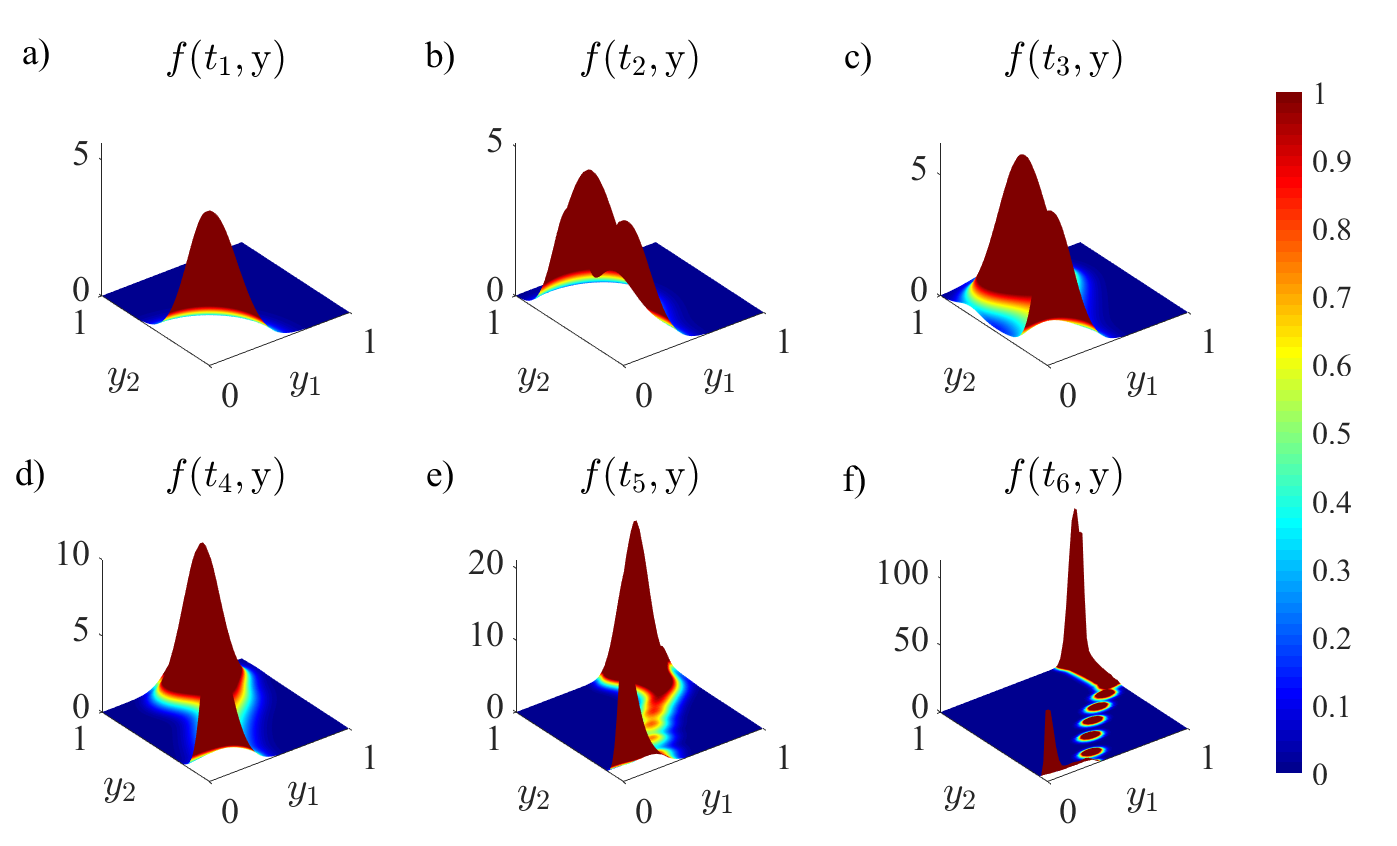}
	\caption{\footnotesize{Plots of the phenotypic distribution of tumour cells across the whole tissue region $f(t,{\bf y})$ defined according to~\eqref{def:f} at six successive time instants $t_1 < t_2 < \ldots < t_6$.}}
	\label{figure7}
\end{figure}

The numerical results of Figure~\ref{figure6} are complemented by the numerical results displayed in Figure~\ref{figure7}, which summarise the time-evolution of the phenotypic distribution of tumour cells across the whole tissue region (\emph{i.e.} the function $f(t,{\bf y})$ defined via~\eqref{def:f}) and show that the maximum point of the distribution departs from the point ${\bf y}=(0,0)$ (\emph{i.e.} the point corresponding to the minimal expression level of both the acidity-resistant gene and the hypoxia-resistant gene) -- \emph{cf.} the initial condition $n^0(x,{\bf y})$ defined via~\eqref{icn} -- and moves toward the point ${\bf y}=(1,1)$, which corresponds to the maximal expression level of both the hypoxia-resistant gene and the acidity-resistant gene (\emph{i.e.} the degree of malignancy of the tumour increases over time). 

\textcolor{black}{These results recapitulate key ideas about the eco-evolutionary dynamics of tumour cells proposed in~\cite{maley2017classifying} by demonstrating how patchy resources (\emph{i.e.} oxygen and glucose) and hazards (\emph{i.e.} the selective pressure effects of lactate), which define the ecology of the tumour, can create multiple habitats whereby different phenotypic variants may be selected according to the principle of the ``survival of the fittest'' (\emph{i.e.} through a functional trade-off between the ability to survive under certain environmental conditions and the evolutionary cost of acquiring such an ability). In the same vein, these results also support the idea that tumour growth can be conceptualised as an ecological process driven by consecutive phases of invasion and colonisation of new tissue habitats by tumour cells, which may be accelerated by the presence of gradients of abiotic factors corresponding to harsher environmental conditions.}

\subsection{Alternative evolutionary pathways leading to the development of resistance to hypoxia and acidity}
\label{evolutionary}
\textcolor{black}{In line with the classification of the evolutionary and ecological features of neoplasms presented in~\cite{maley2017classifying}, the ratio between the selection gradients $\eta_o$ and $\eta_l$ of our model could be seen as a measure of the impact that hypoxia and acidity may have on the eco-evolutionary dynamics of tumour cells. Hence, in this section we explore how the dynamics of the levels of expression of the acidity-resistant gene and the hypoxia-resistant gene across the whole tissue region (\emph{i.e.} the functions $\nu_1(t)$ and $\nu_2(t)$ defined via~\eqref{def:nu}) are affected by the value of the ratio $\eta_o/\eta_l$.}

\begin{figure}[htp!]
		\centering
		\includegraphics[width=0.6\textwidth]{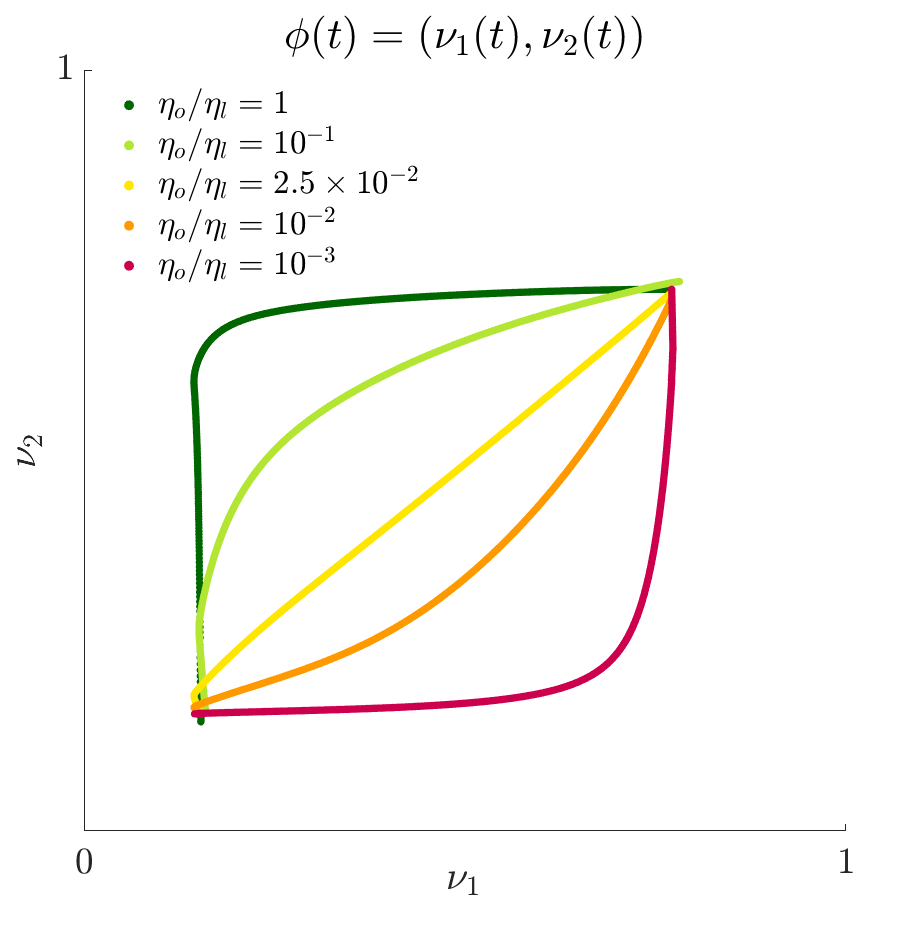}
	\caption{\footnotesize{Plots of the curve $\phi(t) = \big (\nu_1(t), \nu_2(t) \big)$ for  different values of the ratio between the selection gradient related to oxygen $\eta_o$ and the selection gradient related to lactate $\eta_l$. 		The functions $\nu_1(t)$ and $\nu_2(t)$ defined via~\eqref{def:nu} model the levels of expression of the acidity-resistant gene and the hypoxia-resistant gene across the whole tissue region, respectively.}}
		\label{figure8}
	\end{figure}

\textcolor{black}{The plot in Figure~\ref{figure8}, which displays the curve $\phi(t) = \big (\nu_1(t), \nu_2(t) \big)$, demonstrate that resistance to hypoxia and resistance to acidity arise via alternative evolutionary pathways depending on the value of $\eta_o/\eta_l$.} Coherently with the results presented in the previous section, $\phi(t)$ departs from the point $(0.15,0.15)$ (\emph{i.e.} the point corresponding to the initial levels of expression of the acidity-resistant gene and the hypoxia-resistant gene across the whole tissue region) and, for all values of $\eta_o/\eta_l$ considered, ultimately converges to a point corresponding to a high expression level of both the acidity-resistant gene and the hypoxia-resistant gene. However, larger values of the ratio $\eta_o/\eta_l$ lead the level of expression of the hypoxia-resistant gene $\nu_2(t)$ to increase faster than the level of expression of the acidity-resistant gene $\nu_1(t)$, while smaller values of $\eta_o/\eta_l$ correlate with a faster increase of $\nu_1(t)$ and a slower increase of $\nu_2(t)$. Furthermore, for intermediate values of the ratio $\eta_o/\eta_l$ we observe a simultaneous increase of the values of $\nu_1(t)$ and $\nu_2(t)$, whereas sufficiently large and sufficiently small values of $\eta_o/\eta_l$ correlate with a decoupling between the increase of $\nu_1(t)$ and $\nu_2(t)$. In more detail, if the ratio $\eta_o/\eta_l$ is sufficiently high, first $\nu_2(t)$ increases while $\nu_1(t)$ remains almost constant and then, when $\nu_2(t)$ is sufficiently high, $\nu_1(t)$ starts increasing as well. On the other hand, in the case where $\eta_o/\eta_l$ is sufficiently low, we have that $\nu_1(t)$ increases first and then $\nu_2(t)$ starts increasing as soon as $\nu_1(t)$ becomes sufficiently high.

These results communicate the biological notion that: the strength of the selective pressures exerted by oxygen and lactate on tumour cells, which are quantified by the values of the selection gradients $\eta_o$ and $\eta_l$, may shape the emergence of hypoxic resistance and acidic resistance in tumours; the order in which such forms of resistance develop depends on the intensity of oxygen-driven selection in relation to the intensity of lactate-driven selection. \\

\section{Conclusions and research perspectives}
\label{Sec4}
In this work, we have developed a mathematical modelling approach to investigate the influence of hypoxia and acidity on the evolutionary dynamics of cancer cells in vascularised tumours.

The results of numerical simulations of a calibrated version of the model based on real data recapitulate the eco-evolutionary spatial dynamics of tumour cells and their adaptation to hypoxic and acidic microenvironments. In particular, the results obtained indicate that tumour cells characterised by lower levels of expression of hypoxia- and acidity-resistant genes are to be expected to colonise well-oxygenated and mildly-acidic regions of vascularised tissues, whereas cells expressing a more aggressive phenotype characterised by higher levels of resistance to hypoxia and acidity will ultimately populate tissue regions corresponding to hypoxic and acidic microenvironments. Such theoretical findings recapitulate histological data on ductal carcinoma {\it in situ} showing that the levels at which the acidity-resistant gene LAMP2 and the hypoxia-resistant gene GLUT-1 are expressed by cancer cells increase moving from the walls to the centre of the milk duct (\emph{i.e.} moving from more oxygenated and less acidic regions to regions that are less oxygenated and more acidic)~\cite{Damaghi2015ChronicAI,gatenby2007cellular}. 

Moreover, our theoretical findings reconcile the conclusions of~\cite{gatenby2007cellular}, suggesting that tumour cells acquire first resistance to hypoxia and then resistance to acidity, and the conclusions of~\cite{Robertson-Tessi1567}, supporting the idea that the two forms of resistance are acquired in reverse order, by showing that the order in which resistance to hypoxia and resistance to acidity arise depend on the ways in which oxygen and lactate act as environmental stressors in the evolutionary dynamics of tumour cells, which are known to vary between tissue types and between patients~\cite{maley2017classifying}.

We conclude with a brief overview of possible research perspectives. Along the lines of~\cite{lorenzi2020}, the modelling framework presented here could be extended to incorporate additional details of cell movement and mechanical interactions between cells~\cite{ambrosi2002closure,astanin2008multiphase,byrne2003modelling}, which would make it possible to investigate the interplay between phenotypic evolution of cancer cells and tumour growth. \textcolor{black}{Alternative ways of modelling the effect of heritable, spontaneous phenotypic changes, for instance via integral terms~\cite{barles2009concentration,busse2020local,calsina2013asymptotics,diekmann2005dynamics,lorz2013populational}, could also be considered, in order to capture finer details of the processes that drive phenotypic changes~\cite{amadori2015rare,amadori2018rare,carja2017evolutionary,champagnat2006unifying,di2018discrete}. Moreover, it would be relevant to include in the model the effects of stress-induced phenotypic changes caused by hypoxia and acidity, which might be taken into account by introducing a drift term in the balance equation for the local population density function of tumour cells, as similarly done in~\cite{celora2021phenotypic,chisholm2015emergence,lorenzi2015dissecting}. In the vein of \cite{alfaro2019evolutionary,lorenzi2020asymptotic,michod2006life,nguyen2019adaptive}, it would also be interesting to consider possible generalisations of the definition of the fitness function employed here, for instance by letting the selection gradients depend on the concentrations of the abiotic factors, which would allow the concavity of the fitness function to vary across the tumour depending on the local environmental conditions, and by introducing multiple fitness peaks, in order to explore alternative ways in which the spatial and evolutionary dynamics of tumour cells, and their dynamical interactions with abiotic factors, may drive the emergence of intra-tumour phenotypic heterogeneity~\cite{gerlinger2012intratumor}.}

Building on~\cite{ardavseva2019mathematical}, it would be interesting to generalise the model to study the effects of fluctuations in the inflow rate of oxygen and glucose and the outflow rate of lactate in the evolution of cancer cells. In fact, when in hypoxic conditions, cancer cells are known to produce and secrete proangiogenic factors which induce the formation of new blood vessels departing from existing ones. Such an angiogenic process results in the formation of a disordered tumour vasculature whereby the rates at which oxygen and glucose enter the tumour and the rate at which lactate is flushed out through intra-tumour blood vessels fluctuate over time, which impacts on the evolutionary dynamics of cancer cells~\cite{dewhirst2009relationships,kimura1996fluctuations,matsumoto2010imaging,michiels2016cycling}. 

It would also be interesting to extend the model in order to investigate the role of phenotypic transitions triggered by hypoxia and acidity -- such as the epithelial-mesenchymal transition induced by hypoxic environmental conditions~\cite{misra2012hypoxia,tam2020hypoxia,zhang2015hif} and the acquisition of the metastatic phenotype promoted by acidic microenvironments~\cite{Damaghi2015ChronicAI,declerck2010role,fais2014microenvironmental} -- in the phenotypic adaptation of cancer cells and tumour growth.

\textcolor{black}{Furthermore}, since resistance to hypoxia is known to correlate with resistance to chemotherapy and radiotherapy~\cite{cosse2008tumour,declerck2010role,lewin2018evolution,prokopiou2015proliferation,teicher1994hypoxia}, building on~\cite{chaplain2019evolutionary,LORENZI2018101,lorz2015modeling}, it would be relevant for anti-cancer therapy to address numerical optimal control for an extended version of the model that takes into account the effect of chemotherapy and/or radiotherapy~\cite{almeida2019evolution,pouchol2018asymptotic}, which could inform the development of optimised cancer treatment protocols that exploit evolutionary and ecological  principles~\cite{acar2020exploiting,gatenby2009adaptive,korolev2014turning,merlo2006cancer}.\\

\textcolor{black}{Finally, it would certainly be interesting to extend the model presented here to two- and three-dimensional spatial domains. This would make it possible to explore the evolutionary dynamics of tumour cells in a broader range of biological and clinical scenarios and, in particular, it would allow to investigate how the level of tissue vascularisation and the distribution of blood vessels may affect the eco-evolutionary process leading to the emergence of resistance to hypoxia and acidity in vascularised tumours.}
 \section*{Conflict of interest}
The authors declare that they have no conflict of interest.

\appendix
\renewcommand{\thesection}{A.\arabic{section}}

\section{Steady-state properties of the model for the dynamics of tumour cells and robustness of the results obtained}
\label{sec:ApA}
\textcolor{black}{\paragraph{Steady-state properties of the model for the dynamics of tumour cells.} Using a time-scale separation approach and a formal asymptotic method similar to those employed in~\cite{villa2021modeling}, one can formally show that when spontaneous phenotypic changes and undirected, random cell movement occur on slower time scales compared to cell division and death, as in the case of this work (\emph{cf.} the parameter values listed in Table~\ref{tableparameters}), and the fitness function $R(S_o,S_g,S_l,\rho, y_1, y_2)$ is a strictly concave function of ${\bf y}=(y_1,y_2)$ of the form considered here, \emph{i.e.}
\begin{equation}\label{def:Rapp}
R\,(S_o,S_g,S_l,\rho, y_1, y_2) := p\,(S_o,S_g,S_l,y_1, y_2) -  \kappa \rho
\end{equation}
with 
\begin{eqnarray}
\label{def:papp}
p\,(S_o,S_g,S_l,y_1, y_2) &:=& \dfrac{\gamma_o\,S_o}{\alpha_o+S_o}\,\big(1-\varphi_o(S_o)\big) + \dfrac{\gamma_g\,S_g}{\alpha_g+S_g}\,\varphi_o(S_o) \nonumber \\
&& -\eta_o \, \big(y_2-\varphi_o(S_o)\big)^2-\eta_l \, \big(y_1-\varphi_l(S_l)\big)^2
\end{eqnarray}
where $\varphi_o: \mathbb{R}^+ \to [0,1]$ and $\varphi_l: \mathbb{R}^+ \to [0,1]$, the steady-state distribution of tumour cells, $n^{\infty}(x,y_1,y_2)$, will be unimodal and such that the steady-state cell density,
$$
\rho^{\infty}(x) = \int_{0}^1  \int_{0}^1 n^{\infty}(x, y_1, y_2) \, {\rm d}y_1 \, {\rm d}y_2,
$$
the steady-state local mean level of expression of the acidity-resistant gene,
$$
\mu^{\infty}_1(x) =\frac{1}{\rho^{\infty}(x)} \int_{0}^1  \int_{0}^1 y_1 \,n^{\infty}(x, y_1, y_2) \, {\rm d}y_1 \, {\rm d}y_2,
$$
and the steady-state local mean level of expression of the hypoxia-resistant gene,
$$
\mu^{\infty}_2(x) =\frac{1}{\rho^{\infty}(x)} \int_{0}^1  \int_{0}^1 y_2 \,n^{\infty}(x, y_1, y_2)\, {\rm d}y_1 \, {\rm d}y_2,
$$
will satisfy the following system
$$
	\begin{cases}
	R\Big(S_o^{\infty}(x),S_g^{\infty}(x),S_l^{\infty}(x),\rho^{\infty}(x),\mu^{\infty}_1(x), \mu^{\infty}_2(x)\Big)=0,
	\\\\
	\dfrac{\partial R}{\partial y_1}\Big(S_o^{\infty}(x),S_g^{\infty}(x),S_l^{\infty}(x),\rho^{\infty}(x),\mu^{\infty}_1(x), \mu^{\infty}_2(x)\Big)=0,
	\\\\
	\dfrac{\partial R}{\partial y_2}\Big(S_o^{\infty}(x),S_g^{\infty}(x),S_l^{\infty}(x),\rho^{\infty}(x),\mu^{\infty}_1(x), \mu^{\infty}_2(x)\Big)=0,
	\end{cases}
	\quad
	\forall x \in [0,{\rm L}],
$$
where $S_o^{\infty}(x)$, $S_g^{\infty}(x)$ and $S_l^{\infty}(x)$ are the steady-state concentrations of oxygen, glucose and lactate, respectively.}

\textcolor{black}{Substituting definitions~\eqref{def:Rapp} and~\eqref{def:papp} into the above system and solving for $\rho^{\infty}$, $\mu^{\infty}_1$ and $\mu^{\infty}_2$ yields 
\begin{equation}\label{eq:rhoinfty}
\rho^{\infty}(x)= \dfrac{1}{\kappa} \left( \dfrac{\gamma_o\,S_o^{\infty}(x)}{\alpha_o+S_o^{\infty}(x)}\,\big(1-\varphi_o(S_o^{\infty}(x))\big) + \dfrac{\gamma_g\,S_g^{\infty}(x)}{\alpha_g+S_g^{\infty}(x)} \, \varphi_o(S_o^{\infty}(x))\right)
\end{equation}
and
\begin{equation}\label{eq:mu12infty}
\mu^{\infty}_1(x) = \varphi_l(S_l^{\infty}(x)), \quad \mu^{\infty}_2(x) = \varphi_o(S_o^{\infty}(x)).
\end{equation}
}
\textcolor{black}{
These formal results are confirmed both by the plots in Figures~\ref{figure6}c-e, which show that, at every position $x \in [0,{\rm L}]$, the local phenotypic distribution of tumour cells at the end of numerical simulations $n({\rm T},x,{\bf y})$ is unimodal, and by the plots in Figure~\ref{figure9}, which demonstrate that, defining $S_o^{\infty}(x)$, $S_g^{\infty}(x)$ and $S_l^{\infty}(x)$ as $S_o({\rm T}, x)$, $S_g({\rm T}, x)$ and $S_l({\rm T}, x)$ displayed in Figure~\ref{figure4}, respectively, and using the same parameter values as those used to obtain the numerical results of Figures~\ref{figure4} and~\ref{figure6}, there is an excellent quantitative match between: $\rho^{\infty}(x)$ defined via~\eqref{eq:rhoinfty} and $\rho({\rm T},x)$ displayed in Figure~\ref{figure4}; $\mu^{\infty}_1(x)$ defined via~\eqref{eq:mu12infty} and $\mu_1({\rm T},x)$ displayed in Figure~\ref{figure6}; $\mu^{\infty}_2(x)$ defined via~\eqref{eq:mu12infty} and $\mu_2({\rm T},x)$ displayed in Figure~\ref{figure6}.
	\begin{figure}[htp!]
	\centering
	\includegraphics[width=1.02\textwidth]{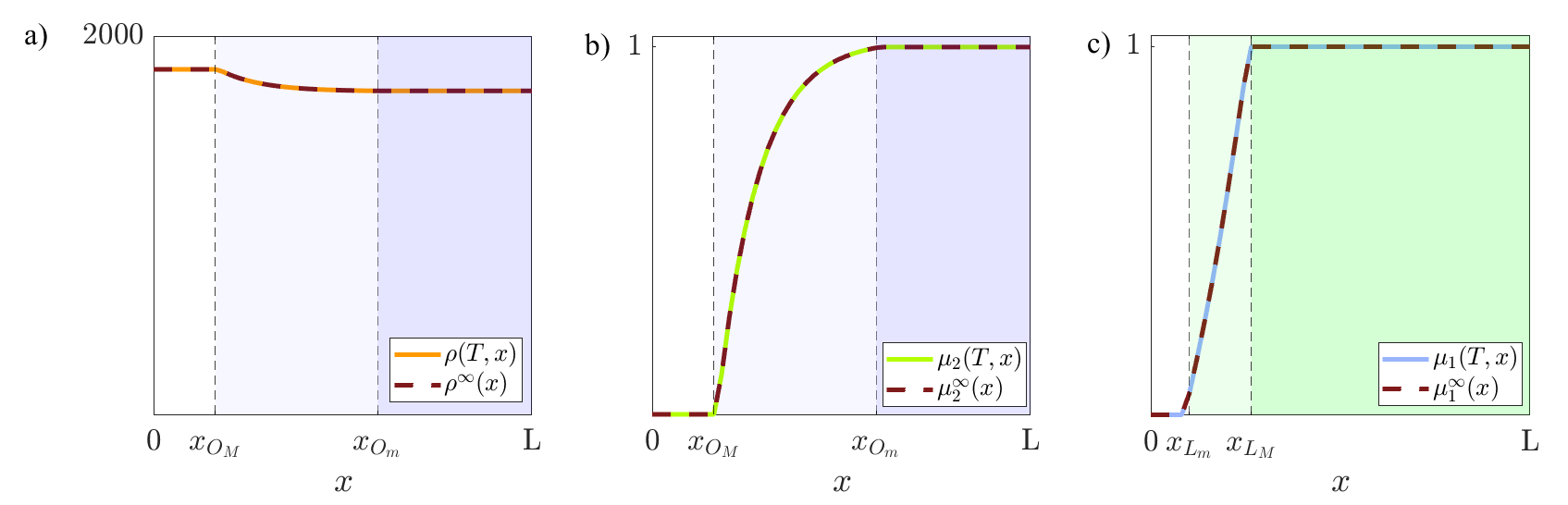}
	\caption{\footnotesize{\textcolor{black}{{\bf a)} Comparison between $\rho^{\infty}(x)$ defined via~\eqref{eq:rhoinfty} (dashed line) and $\rho({\rm T},x)$ displayed in Figure~\ref{figure4} (solid line). {\bf b)} Comparison between $\mu^{\infty}_2(x)$ defined via~\eqref{eq:mu12infty} (dashed line) and $\mu_2({\rm T},x)$ displayed in Figure~\ref{figure6} (solid line). {\bf c)} Comparison between $\mu^{\infty}_1(x)$ defined via~\eqref{eq:mu12infty} (dashed line) and $\mu_1({\rm T},x)$ displayed in Figure~\ref{figure6} (solid line). Here, $\rho^{\infty}(x)$, $\mu^{\infty}_1(x)$ and $\mu^{\infty}_2(x)$ are computed defining $S_o^{\infty}(x)$, $S_g^{\infty}(x)$ and $S_l^{\infty}(x)$ as $S_o({\rm T}, x)$, $S_g({\rm T}, x)$ and $S_l({\rm T}, x)$ displayed in Figure~\ref{figure4}, and using the same parameter values as those used to obtain the numerical results of Figures~\ref{figure4} and~\ref{figure6} (\emph{cf.} Table~\ref{tableparameters}).}}}
\label{figure9}
\end{figure}
}

\textcolor{black}{\paragraph{Robustness of the results obtained.} From the form of the steady-state cell density $\rho^{\infty}$ given by~\eqref{eq:rhoinfty}, the form of the steady-state local mean level of expression of the acidity-resistant gene $\mu^{\infty}_1$ given by~\eqref{eq:mu12infty}, and the form of the steady-state local mean level of expression of the hypoxia-resistant gene $\mu^{\infty}_2$ given by~\eqref{eq:mu12infty}, one can see that the qualitative properties of the results presented in Section~\ref{tumour growth} and Section~\ref{adaptive dynamics} -- \emph{i.e.} the fact that:
\begin{itemize}
\item[i)] the plateau value of the cell density $\rho$ decreases with the distance from the blood vessel;
\item[ii)]  the local mean level of expression of the acidity-resistant gene $\mu_1$ at equilibrium is the minimal one in mildly-acidic regions, the maximal one in highly-acidic conditions and increases with the lactate concentration in moderately-acidic environments;
\item[iii)] the local mean level of expression of the hypoxia-resistant gene $\mu_2$ at equilibrium is the minimal one in normoxic conditions, the maximal one in hypoxic conditions and increases with the oxygen concentration in moderately-oxygenated environments
\end{itemize}
-- remain intact under a broad range of parameter values.}


\end{document}